\documentclass[twocolumn]{aastex631}

\newcommand{\msun}{M_\odot}
\newcommand{\mdot}{\dot{m}}

\usepackage{xcolor}

\usepackage{amsmath}
\usepackage{graphicx}
\usepackage{booktabs}
\graphicspath{figures}
\usepackage{tabularray}

\begin{document}

\title{Thermal and non-thermal emission from supermassive black hole circumbinary disks: \\ Disks, Coronae, Streams, and Cavities}

\author[0000-0003-0685-3525]{Chris Nagele}
\affiliation{Department of Physics and Astronomy\\
Johns Hopkins University\\
Baltimore MD 21218}
\email{chrisnagele.astro@gmail.com}

\author[0000-0002-2995-7717]{Julian H. Krolik}
\affiliation{Department of Physics and Astronomy\\
Johns Hopkins University\\
Baltimore MD 21218}

\author[0000-0002-8676-425X]{Brooks E. Kinch}
\affiliation{Department of Mechanical Engineering and Applied Mechanics\\
University of Pennsylvania\\
Philadelphia PA 19104}

\author[0000-0002-2942-8399]{Jeremy D. Schnittman }
\affiliation{Gravitational Astrophysics Lab, NASA Goddard Space Flight Center, Greenbelt MD 20771}

\author[0000-0003-3547-8306]{Scott C. Noble}
\affiliation{Gravitational Astrophysics Lab, NASA Goddard Space Flight Center, Greenbelt MD 20771}

\begin{abstract}

The search for electromagnetic signals from supermassive black hole (SMBH) binary systems is one of the cornerstones of multi-messenger astrophysics, complementing gravitational wave observations of such systems by pulsar timing arrays and LISA. Although extensive 
simulations have been run to understand the time variability of the bolometric luminosity from accreting binary SMBH systems, comparatively few spectral predictions have been made, and none that go beyond simple emission models. In this paper, we post-process a \texttt{HARM3D} simulation snapshot of a binary at $20 M$ separation accreting at 0.01 Eddington.  For black hole masses $10^6$, $10^7$, and $10^8\, M_\odot$, we
self-consistently solve for the radiated spectrum on the basis of time-steady radiation transfer, thermal balance, and ionization equilibrium, including all relevant relativistic effects as well as emission and absorption processes. Although most of the bolometric luminosity is radiated thermally by 
the disk, the low density regions evacuated by the binary's quadrupole moment produce copious X-rays, with $\sim40\%$ of the observed 
luminosity in a soft X-ray
power law ($\Gamma = 2.3$). We identify two modes of observed azimuthal variation. The X-ray continuum varies by $\sim10\%$ due to an underlying asymmetry in the gas temperature of spiral shocks in the disk. The Fe~K$\alpha$ equivalent width dips by $\sim25\%$ when the line of sight to the inner disk is partially obscured by the lump. These two effects share the same period and are $\sim \pi/2$
out of phase; the
period is order days to weeks for typical AGN masses and a 20$M$ separation.

\end{abstract}


\section{Introduction} \label{sec:intro}


Multi-messenger events, those events where two or more of photons, neutrinos, cosmic rays, and gravitational waves are observed, have tremendous scientific potential.
The presence of multiple messengers enables checks and validations impossible to make with more homogeneous data. 
We are fortunate to have had two robust multi-messenger 
events, photon and neutrino observations of SN1987A \citep{Hirata1987PhRvL..58.1490H} and photon and gravitational wave observations of GW170817 \citep{Abbott2017ApJ...848L..12A}. The former revolutionized the study of massive stars and supernovae \citep{Arnett1989ARA&A..27..629A}. The latter fundamentally changed our understanding of neutron stars, short gamma-ray bursts, and r-process nucleosynthesis \citep{Cowperthwaite2017ApJ...848L..17C}.

One of the exciting frontiers of multi-messenger astrophysics is the low frequency gravitational wave domain, in which supermassive black hole (SMBH) binaries emit gravitational waves. Pulsar timing arrays such as NANOGrav have already observed a stochastic gravitational wave background 
that is likely produced by SMBH binaries \citep{Agazie2023ApJ...952L..37A}.
It may be possible to observe single targets if those targets are close enough or if their locations have been otherwise determined \citep{Agarwal2026ApJ...998L..11A,Veronesi2026arXiv260600218V}. In the coming decade, the Laser Interferometer Space Antenna (LISA) 
will be able to
observe gravitational waves from SMBH binary coalescences, even at high redshift \citep{Amaro-Seoane2023LRR....26....2A}. 

If these gravitational wave observations are to become multi-messenger observations, specific electromagnetic signatures of SMBH binaries must be 
identified because the angular error boxes for gravitational wave detections will contain large numbers of galaxies.  In particular, they must be distinguished from accreting single SMBHs (active galactic nuclei, or AGN).
Many have assumed that the binary's orbital motion would make their most distinctive property a periodic modulation of the light emitted by either
the accretion disks or the jets of the binary's accreting SMBHs.
\citep{Graham2015Natur.518...74G,Liu2016ApJ...833....6L,Charisi2016MNRAS.463.2145C,Liu2019ApJ...884...36L,Chen2020MNRAS.499.2245C,Liao2021MNRAS.500.4025L,Chen2024MNRAS.52712154C,Luo2025ApJ...978...86L}.
Searches for such periodicities have yielded many candidate systems, but none have so far displayed robust evidence for periodic oscillation. 
This exercise is difficult because AGN are red noise dominated \citep{McHardy2004MNRAS.348..783M,Noble2009ApJ...703..964N,Reynolds2009ApJ...692..869R,Kelly2009ApJ...698..895K,MacLeod2010ApJ...721.1014M} 
and can thus masquerade as periodic systems \citep[e.g.][]{Vaughan2016MNRAS.461.3145V}, yet it is worthwhile because such an observation would be transformative.  

A multi-messenger observation of a SMBH binary would provide several immediate benefits \citep{Bogdanovic2022LRR....25....3B}. It would uniquely identify the electromagnetic component as originating from a binary instead of a single SMBH, thereby refining subsequent searches for electromagnetic signatures. It would also provide a robust value for the system's distance (assuming the host galaxy can be identified) and total mass, allowing more precise constraints to be put onto the parameters inferred from the electromagnetic observations. 
If precise enough distances could be determined, these events could be used as standard sirens in cosmology \citep{Abbott2017Natur.551...85A}. Even obtaining separately the individual messenger portions of a multi-messenger event could have great value.  For example, if the EM signature of a SMBH binary were detected, it would confirm that such systems existed.   Numerous such examples could help to define the galaxy population in which SMBHs are formed.

Viscous hydrodynamic simulations \citep{Artymowicz1996ApJ...467L..77A,MacFadyen2008ApJ...672...83M,DOrazio2013MNRAS.436.2997D,Tiede2020ApJ...900...43T,Zrake2021ApJ...909L..13Z} and magnetohydrodymanic simulations \citep{Noble2012ApJ...755...51N,Farris2014ApJ...783..134F,Gold2014PhRvD..89f4060G,Zilhao2015PhRvD..91b4034Z,Bowen2017ApJ...838...42B,Bowen2019ApJ...879...76B,Noble2021ApJ...922..175N,Paschalidis2021ApJ...910L..26P,Bright2023MNRAS.520..392B,Most2024ApJ...973L..19M,Tiwari2025ApJ...986..158T,Manikantan2025PhRvD.112d3004M,Manikantan2025PhRvD.112j3050M,Most2025PhRvD.111h1304M,Ennoggi2026PhRvL.136k1401E}  
of accretion onto a SMBH binary have revealed interesting phenomenology. The binary's quadrupole moment carves out a cavity extending to a few times the binary separation. Inside this cavity, two `minidisks' feed each black hole, and the minidisks are themselves fed by streams emerging from the edge of the cavity. The disk around the cavity is called the circumbinary disk (CBD), and this circumbinary disk may display a notable asymmetry, an $m=1$ (describing azimuthal dependence as $\sum_m a_m e^{im\phi}$) over-density known as the `lump.'
However, the range of conditions in which this lump forms is known only tentatively \citep[][Kim et al. 2026 in prep.]{Noble2021ApJ...922..175N,Tiwari2025ApJ...986..158T}.

In this paper, we focus on the spectral signal from a SMBH binary. We post-process a snapshot from the GRMHD simulations reported in \citet{Noble2021ApJ...922..175N}, which included an equilibrated circumbinary disk as well as the accretion streams, but not the minidisks around each individual black hole. For this simulation snapshot, we solve the spectral radiation transfer problem with all relevant relativistic effects and radiative processes included. Realistic radiation transfer in this regime requires solving not only for thermal balance in each cell of the simulation, but also for ionization balance (involving potentially millions of atomic transitions), in order to properly model the atomic emission and absorption processes within the accretion disk.

We have tackled this problem for single SMBHs \citep{Nagele+2026a,Nagele+2026b}; here we extend that procedure to binaries. 
Our strategy is to divide the simulation into two domains: an optically thin domain wherein we use a relativistic Monte Carlo Compton radiation transfer solver \citep{Schnittman_2013b}; and an optically thick domain, in which we compute a large number of 1D radiation transfer solutions in the fluid's rest frame which include atomic line emission and absorption, as well as Compton scattering and bremsstrahlung \citep{Kinch_2019,Kinch_2021}. These two flavors of radiation transfer calculations interface with each other at the disk's photosphere, each providing the radiative boundary conditions for the other.  By iterating from one to the other, we arrive at a converged radiation field self-consistent over the entire simulation domain. From this final state, we extract detailed spectral predictions for distant observers.

We
summarize our methods in Sec.~\ref{sec:methods}:
the 3D~GRMHD simulation in Sec. \ref{sec:methods_harm}; the Monte Carlo calculation in Sec. \ref{sec:methods_pan}; and the 1D radiation transfer calculations in Sec. \ref{sec:methods_ptx}. Our global iteration is laid out in Sec.~\ref{sec:methods_grand}. In Sec.~\ref{sec:methods_div}, we describe how we divided up the simulation snapshot in order to better understand the complex phenomenology at play. Our results can be found in Sec.~\ref{sec:results}.
In Sec.~\ref{sec:results_thermo}, we describe the thermodynamics of the optically thick and optically thin regions. In Sec.~\ref{sec:results_fluid_spectra}, we describe the nature of the spectral radiation in the local frame of the disk. 
In Sec.~\ref{sec:results_observed} we present the principal result of our calculations, the spectra seen by distant observers. We wrap up with discussion of our results' implications in Sec.~\ref{sec:discussion} and a summary of our conclusions in Sec.~\ref{sec:conclusion}.

\section{Methods} \label{sec:methods}

\begin{figure*}
\includegraphics[width=\linewidth]{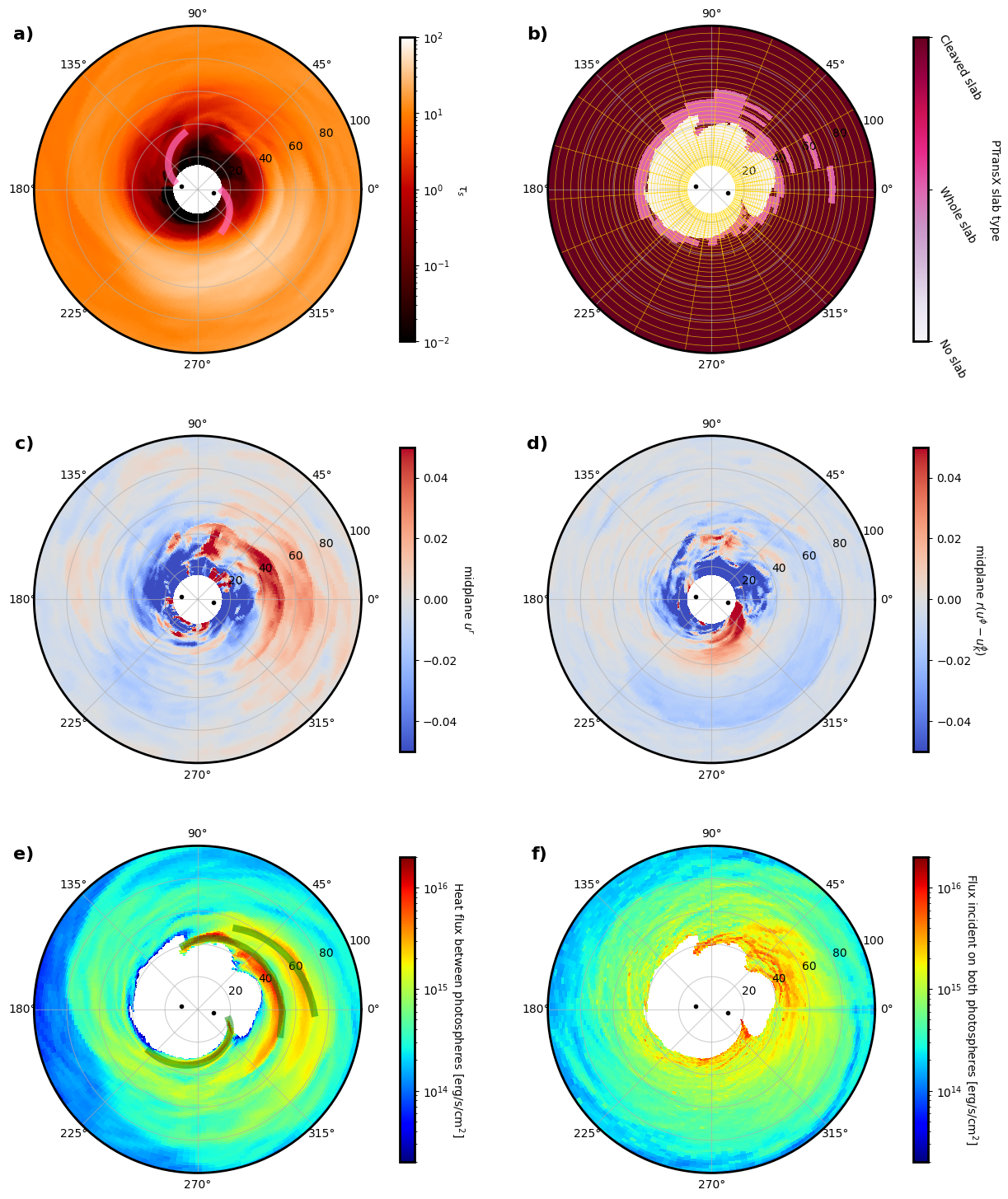}
    \caption{Several quantities in the $r,\phi$ plane for $M=10^6\;\msun$. Scattering optical depth (panel a), type of \texttt{PTransX} slab in that column (panel b), midplane radial velocity (panel c), midplane angular velocity relative to Keplerian rotation (panel d), integrated heat flux between the photospheres (panel e), and integrated flux onto the photosphere (panel f). In panel b, the gold overlay shows the \texttt{PTransX} grid, which coarsens with increasing radius. In panel a, the blue lines show the midplane locations of the streams.
    In panel e, pink lines mark spiral shocks.
    }
    \label{fig:top_down}
\end{figure*}

\subsection{\texttt{HARM3D}} \label{sec:methods_harm}

\texttt{HARM3D} solves the GRMHD equations in three dimensions in flux conservative form \citep{Noble_2009}. In simulations involving only the CBD and the cavity, a post-Newtonian metric of order 2.5 can be utilized, as described in \citet{Mundim2014PhRvD..89h4008M}. Specifically, we make use of the near-zone approximation, which is valid outside $r=0.75\;a$, where $a=20\;r_g$ is the circular-orbit binary separation considered here.  The units of distance and time are $r_g\equiv GM/c^2$ and $r_g/c$ for gravitational constant $G$, speed of light $c$, and total mass of the two black holes $M$. To use this approximation, we 
impose a cutout with a boundary at $r_{\rm in}=15\;r_g$, through which matter and magnetic flux are 
lost from the system. From this inner boundary, the grid extends out to $r_{\rm out}=1000\;r_g$, 
comprising 420 logarithmically spaced radial zones. The simulation uses spherical coordinates, with $N_\theta=160$ and $N_\phi=400$.  The $\theta$-grid concentrates its cells near the midplane \citep{Noble2012ApJ...755...51N}; the $\phi$-grid has uniform spacing in $2\pi$.

For this run and other binary runs, the initial matter distribution is hydrostatic 
with respect to the time average of the run's spacetime \citep{Noble2012ApJ...755...51N}.  Specific solutions depend on the position of the inner edge of the disk, the radius of the pressure maximum, and the aspect ratio $H/r$ at the pressure maximum, where $H$ is the density scale height.  We use $r_{\rm in}=3a$, $r_{\rm P} = 6.5a$ and $H/r=0.1$. The initial magnetic field is a set of nested poloidal loops, normalized so that the volume-integrated ratio of internal energy to magnetic energy is 100. 

The entropy proxy $K$ (defined as $P/\rho^\Gamma$ where $\Gamma=5/3$ is the adiabatic index of the ideal-gas equation of state) is initially set to $K_0 = 0.01$ everywhere. If, during the simulation, the entropy of a fluid element rises above this value, that element cools on a dynamical timescale:
\begin{equation}
    \mathcal{L} = \frac{\rho\epsilon}{t_{\rm cool}}\bigg( \frac{K-K_0}{K_0} + \bigg|\frac{K-K_0}{K_0}\bigg|\bigg)^{1/2}
\end{equation}
where $\rho$ and $\epsilon$ are the fluid rest-frame mass density and internal energy density, and $t_{\rm cool}=2\pi (r/r_g)^{3/2}$ is the period of a circular orbit at radius $r$ for a Schwarzschild black hole of mass $M$. Note that only bound material is cooled.  Unbound material is not cooled for two reasons: it is generally very hot and very low density, potentially invalidating our assumption of a single-temperature thermal gas, and it is almost always found in a high-speed outflow along the polar axis, where it interacts little with any other gas while quickly leaving the system.  
We treat $\mathcal{L}$ as the material's bolometric emissivity in the fluid's rest frame (a Roman $L$ denotes an observed luminosity, as opposed to the fluid frame emissivity), and it, along with $\rho$ and $u^\mu$, the fluid's 4-velocity, are the input data for the post-processed radiation transfer solution.

In this paper, we post-process a single snapshot from the Run$_\mathrm{lrg}$ simulation of \citet{Noble2021ApJ...922..175N}. We choose a snapshot at $t=300000 \; r_g/c$, a time late enough for the disk to have reached inflow equilibrium out to $r=100\;r_g = 5a$. For the post-processing, we include material only out to a radius $r_{L}=440\;r_g$, defined 
so that the post-processing domain contains $99\%$ of the fluid frame luminosity. 
In between 100 and $440 \;r_g$, we use thermal seed photons with a temperature set by the classical disk temperature modulated by the ratio of the cooling in the disk versus the corona, as in \citet{Nagele+2026a}.
In order for the simulation data to be numerically tractable for the post-processing, we must coarsen the simulation snapshot. We linearly interpolate 
the density, emissivity, and velocities first in the $r$ direction and then in the $\phi$ direction, reducing the number of cells by about $40\%$ and $25\%$ respectively \citep{Nagele+2026a}. 

Finally, we convert these quantities to cgs units by assuming an accretion rate $\mdot=0.01$ (in Eddington units), a total black hole mass $M\in 10^{6,7,8}\;\msun$ and a radiative efficiency $\eta=0.0572$ \citep{Novikov1973blho.conf..343N}:
 \begin{align}
\rho_{\text {cgs }} r_g &= \rho_{\text {code }} \frac{4 \pi }{\kappa_T} \frac{\mdot / \eta}{\mdot_{\text {code }}} \label{Eq:rho_cgs}\\
\mathcal{L}_{\text {cgs }} &=\mathcal{L}_{\text {code }} \frac{4 \pi c^3}{\kappa r_g^2} \frac{\mdot / \eta}{\mdot_{\text {code }}}.
\end{align}
where $\kappa_T=0.4$ is the Thompson opacity. We have assumed this relatively low accretion rate because the combination of the relatively large disk scale height (H/r=0.1) with the stipulation that unbound material does not cool means that for larger accretion rates, the photosphere lies in regions of zero cooling, which does not make physical sense.
We have chosen these three masses to explore how our results scale with $M$.
Still larger masses are also of interest, but pose technical difficulties due to the low temperatures deep within their accretion disks \citep{Nagele+2026a}.

\subsection{\texttt{Pandurata}} \label{sec:methods_pan}

\texttt{Pandurata} is a relativistic Monte Carlo code designed for use around Kerr black holes \citep{Schnittman_2013}. Spectral photon packets, which can be thought of as ensembles of photons with different energies, are launched from each patch of the disk-body's photosphere. We define the photosphere as the surface where the scattering optical depth $\tau_s=1$ when the opacity is integrated along a polar angle coordinate curve  
from the nearer polar axis toward 
the midplane. The packets travel along geodesics in a Schwarzschild spacetime 
with a mass $M$ at its center.
We ignore the time-dependence of the spacetime in our geodesic-tracing because the photon travel times are $\lesssim 10^{-3}\times$ the binary's orbital period.  We also ignore 
the binary's quadrupole moment because it does not significantly affect photons near the CBD.

Each cell a packet travels through has a density taken from the GRMHD simulation snapshot (Sec. \ref{sec:methods_harm}); we draw a random number to determine if the packet scatters in that cell. If it does, we draw another random number for the new direction of travel and update the intensity
by the amount of Compton amplification indicated by the 
cell's electron temperature. To do this,
we pre-compute the energy exchange for a large number of Klein-Nishina scattering events as a function of
photon energy and electron energy, weighted according to 
a Maxwell-Juttner distribution in the fluid's rest frame (the cross section is evaluated in the individual electron's rest frame). 
We launch $\sim10^7$ photon packets for each \texttt{Pandurata} run, but this does not give sufficiently good photon statistics in some low density cells, so we utilize the sector scheme as in \citet{Kinch_2019,Kinch_2021,Liu+2025,Nagele+2026a}, with sectors spanning 2x2x2 cells. 

As well as scattering, the photon packet can also reflect off of the disk. To evaluate the results of this process, we make 
use of another pre-computed table, one in which the reflection fraction as a function of photon energy is computed from
a separate Monte Carlo calculation done for each patch of the disk \citep{Kinch_2019}. The inputs to this Monte Carlo calculation are the results of \texttt{PTransX} calculations described in the next section:
the gas temperature, emissivity, and scattering and absorption coefficients as a function of optical depth beneath the photosphere.
Whenever a photon packet reflects off of a photospheric patch, its incoming intensity is recorded for use in the work described in the next section, and its reflected intensity returns to the Monte Carlo transfer solution.
All photon packets eventually reach the inner or outer boundary of the simulation. We assume that photons traveling through the inner boundary are lost to the black holes or minidisks, but we record the direction and spectral intensity of all photons traveling through the outer boundary. 

The initial Monte Carlo solution made by \texttt{Pandurata} assumes that the electron temperature throughout the corona is $T_{e,0} = 50$~keV.  The actual electron temperature 
in thermal balance is defined by equating the inverse Compton power in each sector to the local heating rate found in the underlying simulation.  To solve this equation and determine the temperature, we employ a Newton-Raphson method.  After each \texttt{Pandurata} run, we use the partial derivative with respect to temperature of the inverse Compton power to estimate a new temperature map.  After typically 5-20 iterations, we have a converged solution.

\subsection{\texttt{PTransX}} \label{sec:methods_ptx}

Parallel Transport with XSTAR (\texttt{PTransX}, \citealt{Kinch_2019}) finds the spectrum of light emitted from the photosphere of an accretion disk. 
To do so, at each photospheric location it solves a 1D radiation transfer equation via the Feautrier method \citep{Mihalas1985JCoPh..57....1M}, constrained by local 
thermal and ionization balance. The radiative processes included are: Compton scattering with electron-photon energy exchange and assuming dipolar redistribution; bremsstrahlung; radiative and collisional bound-bound and bound-free transitions; and the radiative emission and absorption rates due to those transitions. To compute these latter quantities, we use routines taken from \texttt{XSTAR}, which solves for atomic ionization states using a large database of reaction rates as functions of 
density, temperature, and incident spectral intensity \citep{Kallman2001ApJS..133..221K}. For a given radiative boundary condition (specified by \texttt{Pandurata}, previous section), the only unknown is thus the gas temperature. Here, too, we use a Newton-Raphson iteration to solve for temperature by cycling through calculations of the radiation and the matter interaction terms until convergence criteria are met (outlined below).

The validity of solving many separate 1D transfer equations rather than a single 3D equation rests upon the assumption
that the radiation intensities vary primarily in the vertical direction. In most locations in our simulation, this is a good assumption. The exceptions are at the inner edge of the accretion disk, where the photosphere is not necessarily normal to the $\theta$ direction. For these 
columns, our solution maintains energy conservation, but does not
necessarily yield the correct
angular distribution of the intensities.  

For each patch of the disk photosphere, we approximate columns of constant $r,\phi$ as being vertical. Each column is discretized with 20-40 grid-points between the upper and lower photospheres, equally spaced in $\log \tau_s$, where $\tau_s$ is the scattering optical depth. 
As for all atmosphere problems, the radiation transfer equation becomes intractable in regions that are close to local thermodynamic equilibrium (LTE), i.e., where the radiation and gas temperatures are nearly equal.
In some columns, this condition is never reached anywhere,
whereas in others the midplane region is dense enough or cool enough to allow thermalization.  We do not know \textit{a priori} whether or where a thermalization surface exists
because thermal equilibrium is achieved through energy exchange, and absorption opacity, along with its inverse, the true emissivity,
depends strongly on the temperature. We 
therefore begin the calculations in all columns assuming they do not have a thermalized core, but if during the temperature iteration, a thermalized core appears,\footnote{Following \citealt{Nagele+2026a}, our criterion for the existence of a thermalized core is that the effective thermalization optical depth $\tau_{\rm therm}>10$ (cf. \citealt{Nagele+2026a}).}
we halt the iteration and split the column in two. We have experimented with pushing the slabs further into the disk body by using a larger thermalization optical depth threshold, but we find that this has no discernible effect on the outgoing spectrum.

We will refer to a single contiguous \texttt{PTransX} domain as a \textit{slab}. A \textit{whole slab} 
is a column without a thermalized core;
a column with a thermalized core
contains two \textit{cleaved slabs}, one adjacent to each photosphere. For cleaved slabs, 
the radiative boundary condition at the thermalization surface is that 
the net flux across the boundary matches half of the dissipation within the thermalized region, so that treated as a whole, the column maintains energy conservation. When iterating the temperature in cleaved slabs, the thermalization surface may be adjusted outward or inward under certain conditions \citep{Nagele+2026a}. 

For each slab, there are two convergence criteria. The first is a test of energy conservation: the net rate of energy gain or loss in the slab 
(i.e., incident plus local heating minus outgoing radiation) must be less than $5\%$ the sum of the integrated heating and cooling rates.
The second is that for successive temperature iterations that satisfy the energy conservation criterion, the outgoing spectral flux must not differ for the two iterations by more than $5\%$ in each energy bin
within one decade in energy of the peak of the outgoing flux.

Because it solves for ionization balance, \texttt{PTransX} is too expensive to run on every column of the simulation. 
Much as we do for \texttt{Pandurata}, we therefore selectively coarsen the grid, as illustrated
by the gold overlay in Fig.~\ref{fig:top_down}b. The radial resolution of the \texttt{PTransX} grid
is homogeneous (every third \texttt{Pandurata} cell), but
the azimuthal resolution 
is $2/4/8$ times as coarse as the simulation gridscale outside $r = 20/30/40\;r_g$.
This change is demanded in order to capture the energetics at the inner edge of the disk \citep[Appendix A of][]{Nagele+2026a}. After a \texttt{PTransX} calculation, the outgoing flux from successful columns is used to set the seed photon spectrum at each photosphere location for the next \texttt{Pandurata} run  (a few $\%$ of slabs fail to converge, and for these slabs we use the results from the nearest successful column).

\begin{figure*}
\includegraphics[width=\linewidth]{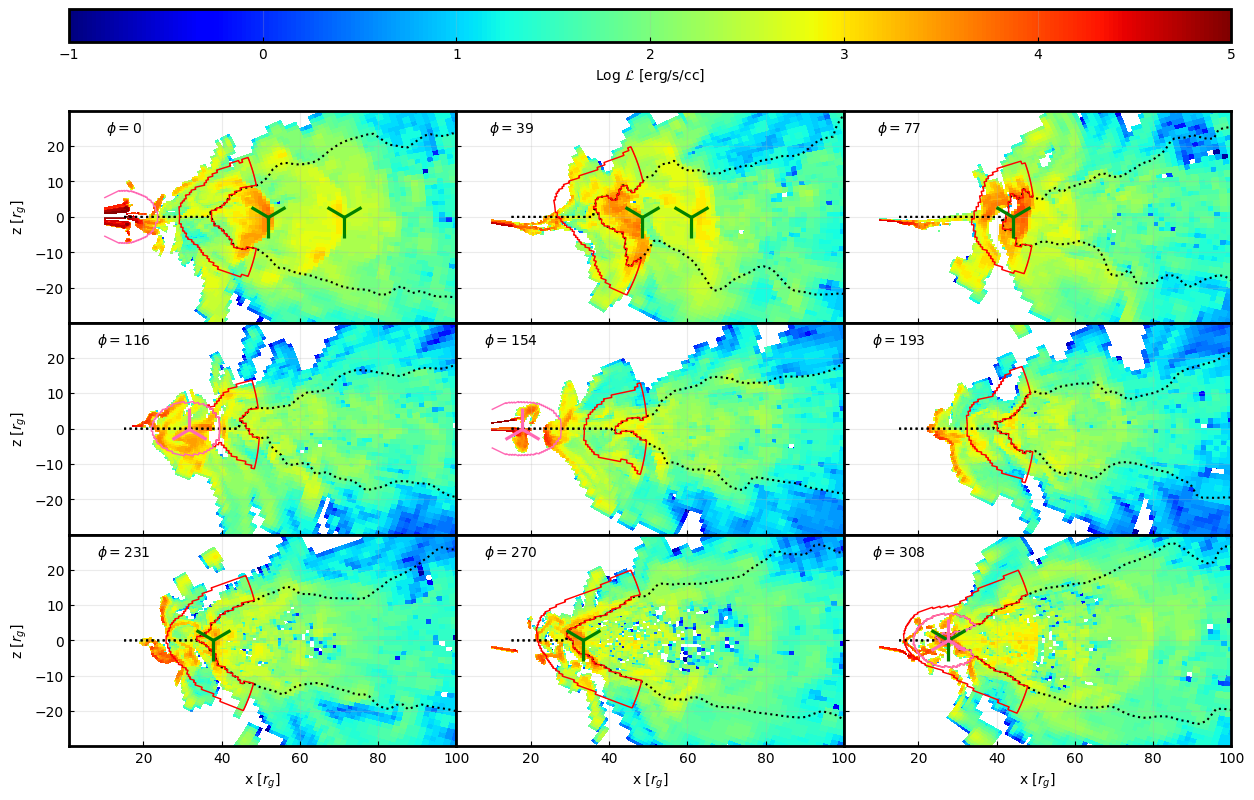}
    \caption{Azimuthal slices of \texttt{HARM3D} luminosity (cooling function) for $M=10^6\;\msun$ on the \texttt{Pandurata/PTransX} grid. In each panel, the black dotted line shows the $\tau=1$ scattering photosphere. White regions are unbound material, which is assumed not to cool. The red contour encloses material associated with the cavity wall. The pink contour (in panels $\phi = 0^\circ, \, 154^\circ, \, 308^\circ$) encloses material associated with the streams. The pink upward-facing triple-prong symbols
    (in panels $\phi = 116^\circ, \, 154^\circ, \, 308^\circ$)
    mark the midplane location of the streams 
    (Fig. \ref{fig:top_down}a).     
    Pink downward-pointing triple-prong symbols
    (in panels $\phi = 0^\circ, \, 39^\circ, \, 77^\circ, \, 231^\circ, \, 270^\circ, \, 308^\circ$)
    mark disk columns associated with the spiral shocks (Fig. \ref{fig:top_down}e). In the final panel of this figure ($\phi=308^\circ$) the stream coincides with a shock. 
     }
    \label{fig:azimuth_cf}
\end{figure*}

\begin{figure*}
\includegraphics[width=\linewidth]{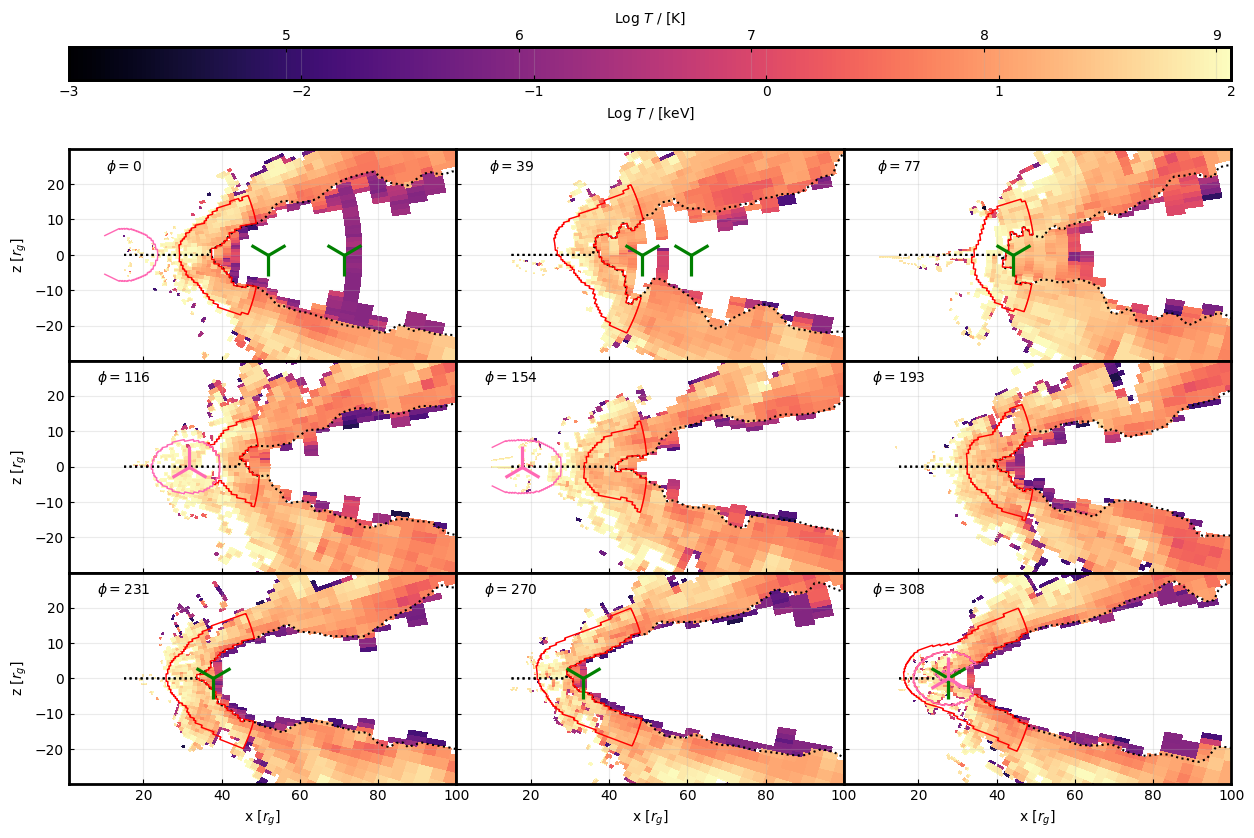}
    \caption{Analogous to Fig. \ref{fig:azimuth_cf}, but showing temperature instead of luminosity. The temperature outside of the scattering photosphere (black dotted line) is determined by \texttt{Pandurata}, and white regions are either a) off-scale, b) have zero \texttt{HARM3D} luminosity or c) have zero \texttt{Pandurata} scatterings, due to low density. Inside the photosphere, temperatures are determined by \texttt{PTransX} and white regions are either the location of the thermal core or slabs that fail to converge (e.g. bottom right of the fifth panel).}
    \label{fig:azimuth_T}
\end{figure*}

\subsection{Grand iterations}\label{sec:methods_grand}

We begin the post-processing with a \texttt{Pandurata} run 
in which
the seed photons are assumed to be a hardened blackbody. The temperature of the blackbody is determined by integrating the heating rate in each column of the disk-body, after which a hardening factor of 1.7 is applied. In general, the seed photon spectrum of this first \texttt{Pandurata} run does 
not resemble the final spectrum, but we 
must begin with some 
assumption.

After the first \texttt{Pandurata} run, the calculations proceed in the following sequence: 
\begin{enumerate}
    \item \texttt{PTransX} with boundary flux from last \texttt{Pandurata} run.
    \item Reflection Monte Carlo with internal disk temperature 
    ionization states from the last \texttt{PTransX} run.
    \item \texttt{Pandurata} with seed photon spectra from last \texttt{PTransX} run and using the updated disk reflection table.
\end{enumerate}
We refer to this as the \textit{grand iteration}. In order for us to deem this whole apparatus converged, solid-angle averaged
spectra from successive \texttt{Pandurata} runs must agree to within $5\%$ for each energy bin in the range 10~eV$ \leq E \leq10$~keV (where $E$ is the photon energy). The cases studied in this paper took 5-6 grand iterations to converge. 

For the bulk of the grand iterations, we use a relatively coarse energy grid, with 20 gridpoints per decade in the range 1~eV $\leq E\leq 10$~MeV. Once the system has converged, however, we perform one additional calculation of both \texttt{PTransX} and \texttt{Pandurata} using a higher resolution energy grid with 400 gridpoints per decade. The continuum spectra from these two calculations agree exactly, but
the higher resolution calculation has much better-resolved atomic emission lines and absorption troughs.
All the spectra we present use the higher energy resolution.

\subsection{Divisions of the simulation}\label{sec:methods_div}

In GRMHD simulations of single black hole accretion, the geometry is roughly axisymmetric, even for a single time snapshot. Around a binary, the situation is more complicated, and we devote this subsection to the description of the different 
subregions found in this \texttt{HARM3D} snapshot. 

Fig.~\ref{fig:top_down} shows several variables as functions of $(r,\phi)$. Panel a shows the scattering optical depth to the disk's midplane. 
We place any column whose midplane $\tau_s > 1$ in the circumbinary {\it disk body}.  Consequently, in
the central white region of Fig.~\ref{fig:top_down}b,
$\tau_s<1$ and 
this region is treated solely by \texttt{Pandurata}. Note that we will often take azimuthal averages of the disk properties, and 
at small radii ($\lesssim 40r_g$), the only places where there is significant gas are in the narrow accretion streams; consequently,
there may be only a few optically thick columns
at $20r_g \lesssim r \lesssim 40r_g$.
The surface density within the disk body varies fairly smoothly, as can be seen in Figure~\ref{fig:top_down}a, but there is a distinct concentration centered on $r\sim50r_g,\phi\sim300^\circ$.  This is the ``lump" identified by \citet{Shi2012ApJ...749..118S} and studied in detail in \citet{Noble2021ApJ...922..175N}.

A second noteworthy region is that occupied by the {\it accretion streams} falling ballistically toward the black hole binary. These can also be seen in Figure~\ref{fig:top_down}a. Blue lines have been overlaid to trace the streams,
defined by power law fits  to the 
azimuthal angle $\phi_{\rm max}$ at which the midplane density at each radius is greatest, i.e., we find $r_s$ and $a$ in the relation $\phi_{\rm max} = r_s r^a$. We define the region occupied by each stream as having a half-width of $7.5r_g$ and extending out to a radius where the radial velocity (3rd panel) switches sign from infalling to outgoing.  The outer end is usually found at $r \approx 35 - 40 r_g$, a slightly smaller radius than the inner edge of the circumbinary disk.   At almost every time, there are two such streams, roughly opposite one another.

In addition to the streams, it is worthwhile to define three more subregions, all of them within the volume where $\tau_s < 1$:
\begin{enumerate}
 \item Locations outside $r =50 r_g = 2.5 a$ are in the \textit{corona}.
    \item Inside $r =50 r_g = 2.5 a$, cells farther than $7.5 r_g$ from the center of a stream but within 7.5 $r_g$ of the disk body photosphere are part of the \textit{cavity wall}.
    \item All other cells inside $r =50 r_g = 2.5 a$ are in the \textit{cavity}.
\end{enumerate}
These criteria define the colored curves in Fig.~\ref{fig:azimuth_cf}.
Although grounded in real physical contrasts, these definitions are  more qualitative than quantitative, and the boundaries they define are, in reality, rather fuzzy. Nonetheless, they do correspond to physical regimes distinct enough to separate.

\section{Results} \label{sec:results}

\subsection{Thermodynamics} \label{sec:results_thermo}

\subsubsection{Heating mechanisms}
\label{sec:results_heating}

As is easily visible in Figs.~\ref{fig:top_down}e and \ref{fig:azimuth_cf}, the heating rate is very strongly concentrated.  This concentration is a consequence of the nature of the dissipation mechanisms.  Two are most important: hydrodynamic shocks and magnetic reconnection.  For both, heating occurs in thin surfaces, hence the concentration.

Fig. \ref{fig:top_down} has two sets of curves overplotted, streams in panel a) and spiral shocks in panel e). The streams are located in regions of high $|\nabla\times \vec B|$. The curl of the magnetic field is often associated with magnetic field dissipation because it is large where oppositely directed field lines are close together \citep{Hirose2006}.  The dissipation in the streams is likely to be mostly magnetic because, in addition to the high current density, in all but a small part of this region the local magnetic pressure is at least an order of magnitude greater than the local gas pressure. 

Similar to the streams, the shocks' ultimate origin is the binary torques.  When matter first falls inward from the inner edge of the CBD, it typically has an angular momentum only slightly smaller than the circular orbit angular momentum at that radius.  It therefore moves inward only slowly and has time to receive additional angular momentum from the binary torques.  Matter with increased angular momentum swings outward, and when it hits the inner edge of the CBD, it launches an outward-moving shock wave whose front is stretched into a spiral by orbital shear.

\begin{figure*}
\centering
\includegraphics[width=\linewidth]{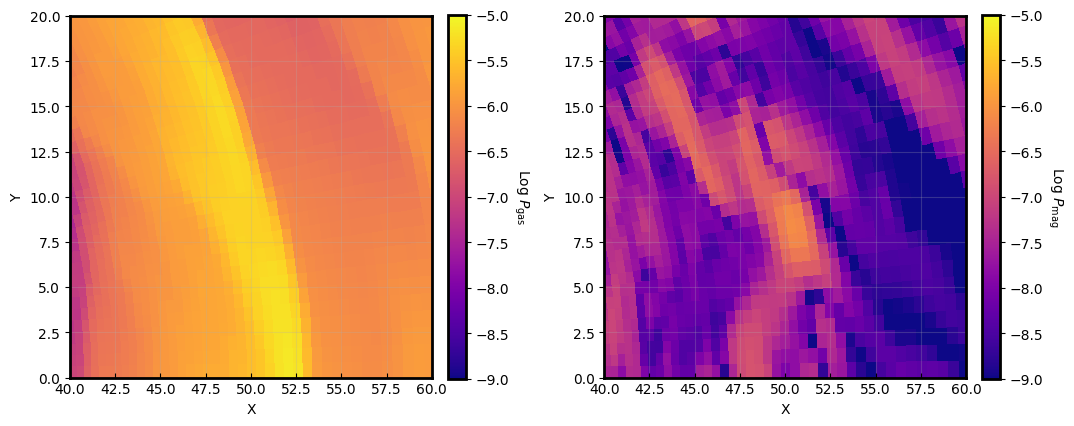}
    \caption{Midplane slices of gas pressure and magnetic pressure, both in code units. This figure uses the native \texttt{HARM3D} resolution and is zoomed in on one of the spiral shocks.}
    \label{fig:curlB}
\end{figure*}

In the spiral shocks, $|\nabla\times \vec B|$ is also high, raising the possibility that the heating in the spiral shock as well is magnetic. To address this question, we show gas pressure and magnetic pressure zoomed in on one of the spiral shocks located at $r=50\;r_g$ in Fig. \ref{fig:curlB}. The edge of the shock is clearly visible in the gas pressure slice, and the shock is moving to the right (outwards in radius, cf. Fig. \ref{fig:top_down}). The magnetic pressure is high, sometimes almost comparable to the gas pressure downstream from the shock (this is the region which has a high $|\nabla\times \vec B|$), but is multiple orders of magnitude lower upstream (where $|\nabla\times \vec B|$ is negligeable). 
Thus, magnetic dissipation may contribute to the heating in the downstream region of the shock because 
post-shock compression has amplified the magnetic field,
but the primary dissipation occurs in the shock front and is hydrodynamic.

\subsubsection{Optically thick regions} \label{sec:results_thick}

The most basic result of our \texttt{PTransX} calculations in the optically thick disk is
the determination of which columns contain a whole slab and which contain two cleaved slabs
(Sec.~\ref{sec:methods_ptx}). For the large radii of the circumbinary disk in this study ($50-100\;r_g$), we would expect the overwhelming majority of columns to have thermal cores, as there is typically not enough internal or external heating to keep the temperature high enough to avoid  thermalization.

That is broadly true in this case, as can be seen in the sample poloidal slices 
shown in Fig.~\ref{fig:azimuth_T}. In the first and third panels, however, there are columns containing whole slabs out to very large radii, including one at $70\;r_g$ in the first panel, and there are numerous whole slabs out to $\approx 50 - 60 r_g$ in the third panel. 
These whole slabs at large radius are found either in regions of strong shock heating or in regions of low density between spiral shocks. Low density leads to high temperature because the incident flux has little spatial variation, so low density promotes high ionization and therefore weak cooling.

In columns located near the spiral shocks, the temperature is high for both whole slabs and cleaved (Fig. \ref{fig:azimuth_T}). The high temperatures in these columns lead to higher ionization states, which suppress
absorption; weaker absorption makes 
photons more likely to be reflected by Compton scattering, in which on average they lose little energy. This means the radiation coming out of these hot, shock heated, slabs 
has a harder spectrum than the spectrum radiated by neighboring slabs that have not been shock heated
(Sec. \ref{sec:results_outgoing}).

Although the disk structure clearly exhibits variation in azimuth, it is still useful to consider the radial dependence of several azimuthally-averaged
disk properties. Fig.~\ref{fig:Tr} shows the radial dependence of several kinds of temperature in the simulation, including the photospheric gas temperature and ``core flux temperature" where a thermal core exists.   We define the core flux temperature as the effective temperature associated with the integrated heating rate inside the core. 
The photosphere temperatures are generally high, similar to the coronal temperatures, and
Fig.~\ref{fig:azimuth_T} shows that in the majority of slabs the gas temperature falls 
rapidly with increasing optical depth, dropping from $\sim 10^7 - 10^8$~K at the photosphere to $\sim 10^5 - 10^6K$ at the thermalization surface.  The lowest ``temperature" is the core flux temperature, which is $\sim 10^4 - 10^5$~K (triangles in Fig. \ref{fig:Tr}).  These trends can be understood as the result of two effects.  At the thermalization surface, by definition, the gas temperature is close to the radiation temperature, but photon scattering outside the thermalization surface holds back photon escape, raising the temperature at that location well above the flux temperature.  Farther from the core, the radiation energy density decreases, but outside the thermalization surface the gas temperature decouples from the radiation temperature and rises outward.

\subsubsection{Optically thin regions} \label{sec:results_thin}

\begin{figure}
\includegraphics[width=\linewidth]{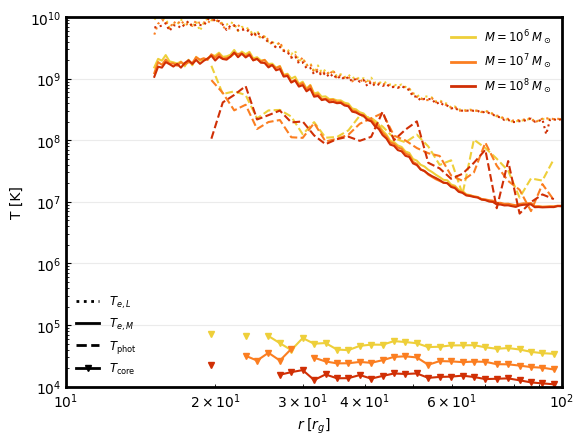}
    \caption{Several
    temperatures as a function of radius.  The curves are: coronal electron temperature weighted by cell luminosity (dotted line, averaged over solid angle), coronal electron temperature weighted by cell mass (solid line, averaged over solid angle), photosphere temperature determined by \texttt{PTransX} (dashed line, averaged over azimuthal angle) and the core flux temperature of cleaved slabs (inverted triangles, averaged over azimuthal angle).  
    }
    \label{fig:Tr}
\end{figure}

In optically thin regions, the temperature is determined by the balance between the heating rate defined by the cooling function \citep{Noble_2009} in the GRMHD simulation \citep{Noble2021ApJ...922..175N} and the inverse Compton power determined by \texttt{Pandurata}. Hotter regions in the corona tend to have some combination of a high intrinsic heating
rate, a low electron density, or a low radiation energy density. The other trend that
can be seen in this figure and in Fig.~\ref{fig:Tr} is that the temperature decreases with increasing radius.
Both the mass-weighted and the luminosity-weighted coronal temperature averages (see Fig.~\ref{fig:Tr}) decrease with increasing radius out to the edge of the \texttt{PTransX} domain at $r=100\;r_g$. 

\subsubsection{The subregions' luminosity shares}

\begin{figure}
\includegraphics[width=\linewidth]{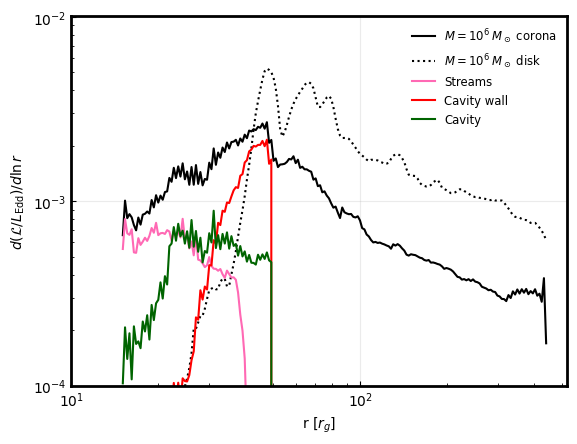}
    \caption{Volume integrated rest frame luminosity in Eddington units. The black dashed line shows the disk-body, while the black solid line shows the corona. The colored lines indicated contributions to the latter quantity from various locations in the corona inside $r=50 \;r_g$. }
    \label{fig:dLdr}
\end{figure}

The total fluid-frame luminosity is shared 59\%/41\% between optically thick and thin regions.
As made clear by Figure~\ref{fig:dLdr}, essentially all the optically thick luminosity is radiated outside $\approx 40 r_g$. In the region $15<r/r_g<20$, the optically thin luminosity is dominated by the streams, while in the region $30<r/r_g<50$, it is dominated by the cavity wall. In between, all three types of optically thin regions meaningfully contribute.

The total optically thin luminosity is likely an underestimate in two ways.
First, this \texttt{HARM3D} simulation used the target-entropy cooling rate of \citet{Noble_2009} rather than the approximate Compton cooling of \citet{Kinch_2020}. Switching from the target entropy cooling function to the Compton cooling version has shown that the former tends to underestimate the cooling in the corona by a factor order unity.
Second, in very low-density regions, 
particle populations given additional energy in shocks, magnetic reconnection, etc.  may reach thermal distributions rather slowly,
significantly complicating any calculation of their emissivity.   Although such complications 
have been parameterized in GRMHD simulations of the low luminosity AGN regime \citep{Pandya2016ApJ...822...34P,EHT2019ApJ...875L...5E,Marszewski2021ApJ...921...17M,EHT2022ApJ...930L..16E} and 
studied directly in particle-in-cell (PIC) simulations of AGN coronae \citep{Groselj2026arXiv260100518G}, many questions remain open.  To avoid these difficulties in the current study, the cooling rate is set to zero in unbound material (most of the white regions in Fig.~\ref{fig:azimuth_cf}), which tend to have especially low densities.

\subsection{Fluid frame spectra} \label{sec:results_fluid_spectra}

\subsubsection{Disk incident flux} \label{sec:results_incident}

\begin{figure*}
 \includegraphics[width=\linewidth]{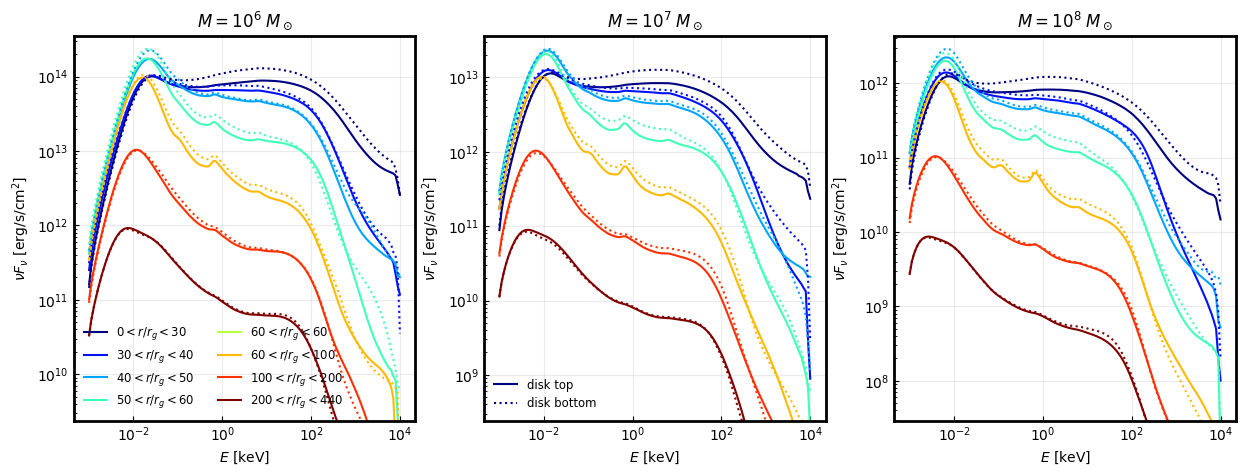}
    \caption{\texttt{Pandurata} spectral flux incident on the photosphere averaged over different radial bins (colors) for both the top (solid line) and bottom (dotted line) photospheres, for each of the three black hole masses (panels). }
    \label{fig:flux_incident}
\end{figure*}

Fig.~\ref{fig:flux_incident} shows the spectrum of the incident flux averaged over the photospheric areas of eight annuli for all three binary masses and the top and bottom surfaces of the disk.
Overall, the flux decreases with increasing radius.  However, its spectrum also changes, softening with increasing radius.
Inside the CBD edge, $\nu F_\nu$ is nearly flat from $\simeq 20$~eV to $\sim 100$~keV
for all three black hole masses.
Outside $r\approx 40\;r_g$, $\nu F_\nu$ is roughly $\propto \nu^{-1}$.

At the lower energy end of the spectrum, a thermal peak is visible. This peak represents
outgoing photon packets whose spectrum is mostly thermal and return to the disk after scattering a single time. For these photons, the corona essentially acts like a mirror, and the spectrum of the flux returning to the disk thus mimics the emitted flux at photon energies where there is little intrinsic coronal emission. The same process imprints emission line features such as Fe K$\alpha$ 
and OVIII Ly$\alpha$
on the spectrum striking the disk.

At the high energy end, the power law breaks around $E=100\;$ keV for the $M=10^6\;\msun$ case. The break is at slightly lower energy for higher masses. There is also an overall softening of the power law with increasing mass.

\begin{figure}
\includegraphics[width=\linewidth]{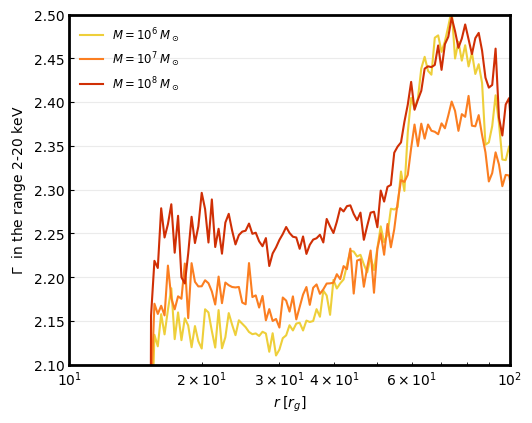}
    \caption{Photon index $\Gamma$ of the $2-20$~keV flux incident on the disk, averaged over top/bottom as well as $\phi$.
    }
    \label{fig:Gamma}
\end{figure}

Both the photon index and the ionization parameter vary with position on the photosphere. 
Note that both of these quantities have substantial azimuthal dependence as well as radial dependence for the reasons we discussed above (cf. Fig. \ref{fig:top_down}e).
For concision, we will describe only the radial dependence and not the azimuthal dependence.
For the photon index, we summarize the radial-dependence by fitting the spectrum on each photospheric patch to a power law $\nu F_\nu \propto \nu^{2-\Gamma}$ 
(Fig.~\ref{fig:Gamma}) and then averaging over azimuth. Note that at some particularly small radii, only a few patches of photosphere exist because most of the material is optically thin (Sec. \ref{sec:methods_div}). For all three black hole masses, 
the flux softens with greater radius out to $r=100\;r_g$, but then hardens again outside this radius, only to soften further at the very edge of the computational domain. This re-hardening is likely an artifact of inflow equilibrium extending to only $r=100\;r_g$.

\begin{figure}
\includegraphics[width=\linewidth]{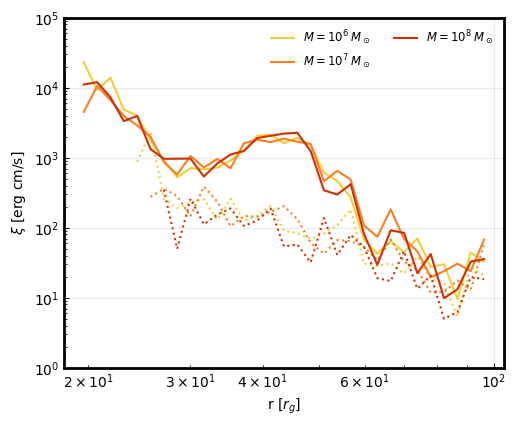}
    \caption{Ionization parameter at the disk surface (solid lines) and the thermal core (dotted lines). As in Fig.~\ref{fig:Tr}, 
    both quantities have been averaged over azimuthal angle.
    }
    \label{fig:Xi}
\end{figure}

Fig.~\ref{fig:Xi} shows the radial dependence of the ionization parameter at the photosphere (solid lines) and at the thermalization photosphere (dotted lines), similarly averaged over azimuth. At the inner edge of the disk, most slabs are in the fully or highly ionized regimes, but outside $r = 50\;r_g$, 
most are in lower ionization states. There does not appear to be any dependence of the ionization parameter on black hole mass.

\subsubsection{Disk outgoing flux} \label{sec:results_outgoing}

\begin{figure}
 \includegraphics[width=\linewidth]{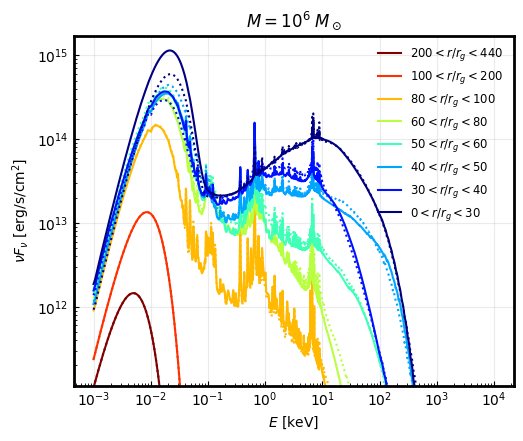}
    \caption{\texttt{PTransX} spectral outgoing photosphere flux (\texttt{Pandurata} seed photons) averaged over different radial bins (colors) for both the top (solid line) and bottom (dotted line) photospheres, for $M=10^6\;\msun$. }
    \label{fig:flux_out}
\end{figure}

The flux traveling out of the disk at the photospheric boundary is computed by \texttt{PTransX} using the incident flux discussed in the previous section as the radiative boundary condition. Fig.~\ref{fig:flux_out} is the figure directly analogous to Fig.~\ref{fig:flux_incident}, showing the spectral flux averaged over different radial bins.  Note that because the average includes only locations where a photosphere exists, for the innermost radii only a handful of columns count toward the average (see Fig.~\ref{fig:top_down}). There are two distinct energy regimes in this figure: for $E>0.1$~keV, the spectrum can be roughly described as a broken power law with considerable flux in emission lines; for $E\leq0.1$~keV, it is dominated by a thermal spectrum. 
As for the flux incident on the disk, the spectral slope at energies above the power law break becomes softer with increasing radius.
Although it is difficult to locate precisely the ``break" between the power laws, it is generally found at $E\approx 10$ keV for $r\leq60\;r_g$. Emission lines are prominent in the spectra of even the highly-ionized innermost annuli (Fig. \ref{fig:Xi}) and grow even further in equivalent width with increasing radius. For the outermost radial bin 
treated
by \texttt{PTransX} ($80<r/r_g<100$), the emission lines are so dense that the continuum level cannot be readily identified. The thermal part of the spectrum decreases in flux and temperature with radius, as the corresponding 
core flux temperature of the disk decreases (triangles in Fig. \ref{fig:Tr}).

\begin{figure*}
\includegraphics[width=\linewidth]{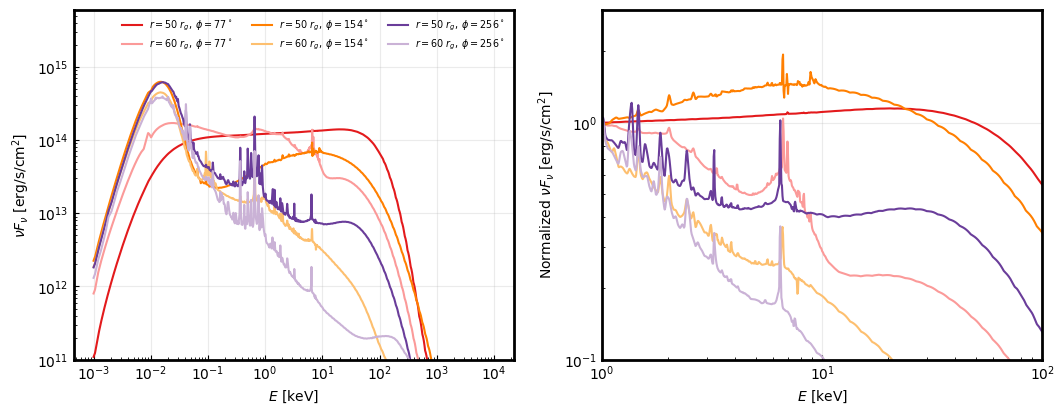}
    \caption{Six examples of the \texttt{PTransX} outgoing flux, at two different radii and three azimuths, corresponding to three optical depth regimes (Fig. \ref{fig:top_down}a). Note that unlike Fig. \ref{fig:flux_out}, this figure is showing the true outgoing flux, instead of the seed flux. This includes the flux 
    reflected during the \texttt{Pandurata} run (cf. \citealt{Kinch_2019}). The right panel shows the same curves, but zoomed in on the X-ray band and normalized at $E=1$ keV for legibility.}
    \label{fig:flux_out_comparison}
\end{figure*}

As well as the radial dependence discussed above, there can also be strong azimuthal variation. 
That this can happen is of intrinsic interest, but it is also a potential source of systematic time-variability, which we will discuss in
Sec.~\ref{sec:results_observed}. Fig.~\ref{fig:flux_out_comparison} shows the outgoing flux for six slabs, sampled at 
two radii ($50,60\;r_g)$ and three azimuths ($\phi=77^\circ,154^\circ,256^\circ$). 
The $r=50\;r_g$, $\phi=77^\circ$ column is very close to a spiral shock, whose strong heating  can produce strong ionization and low opacity that suppresses X-ray absorption so that the emitted X-ray spectrum is harder and there is essentially no thermal component.  The $r=60r_g$, $\phi = 77^\circ$ column is farther from any shock, but is particularly low density, which also supports strong ionization and a similar, although slightly softer, spectrum.
In sharp contrast, at $r=50\;r_g,\phi=154^\circ,256^\circ$ there is neither low densities nor high heating rates.
As a result, most of the luminosity is in a quasi-thermal spectrum and X-ray emission lines are prominent.

The second panel of Fig.~\ref{fig:flux_out_comparison} zooms in on the 
1 - 100~keV band to allow the emission lines
to stand out more clearly.
There is a remarkable range in the character of these features within the narrow annulus $50 r_g \leq r \leq 60r_g$.
The $\phi=77^\circ,\;r=50\;r_g$ spectrum has a barely visible H-like Fe~K$\alpha$ line. This line is so weak due to the highly ionized state of this slab as well as the high level of the X-ray continuum. At $\phi=77^\circ,\;r=60\;r_g$ there is a much stronger and extremely Compton-broadened Fe~K$\alpha$ line underlying the two narrow components at $E=6.7,6.97$~keV for He-like and H-like iron. Both $\phi=154^\circ$ spectra contain emission lines and a few absorption lines, while the $\phi=256^\circ$ spectra, whose slabs sit atop the lump, are dominated by a plethora of emission lines in which Fe~K$\alpha$ comes from moderately-ionized atoms.

\begin{figure}
\includegraphics[width=\linewidth]{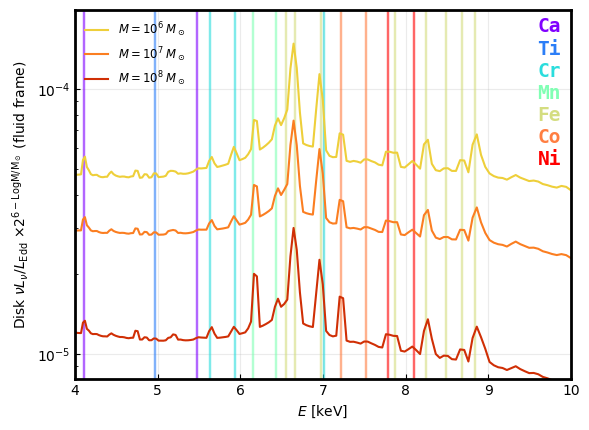}
    \caption{Fluid frame disk spectra in the Fe K$\alpha$ region for each of the three black hole masses. Vertical 
    lines show the elemental provenance of emission lines at those energies. Note that there are two lines very close to each other around 7 keV, a Fe K$\alpha$ line and a Cr K$\beta$ line. }
    \label{fig:Ldisk_Kalpha}
\end{figure}

Finally, in Fig.~\ref{fig:Ldisk_Kalpha}  
we show the seed photon flux in the fluid frame integrated over the disk surface, this time focusing specifically on the Fe~K$\alpha$ region. This figure
shows luminosities rather than fluxes to give a sense of how much the different elements and their various ionization states contribute to emission lines with energies close to Fe~K$\alpha$.
The spectra are dominated by the He- and H-like features at $6.7$ and 6.97 keV, respectively, although some lines from less ionized states can be seen between 6.4 and 6.7 keV. 
In addition, several other iron features can be seen, including Fe~K$\beta$ at 7.8 and 8.2~keV and a bump just above the He-like Fe recombination edge at 8.8~keV. Note that the recombination feature's profile is different from that of
the emission lines. 
Although $\simeq 90\%$ of the line emission is from iron, K$\alpha$ and K$\beta$ lines from other iron peak elements (Cr, Mn, Co, and Ni) can supplement the emission line luminosity in this energy range.

\subsection{Observed spectra} \label{sec:results_observed}

\begin{figure*}
\centering
\includegraphics[width=0.9\linewidth]{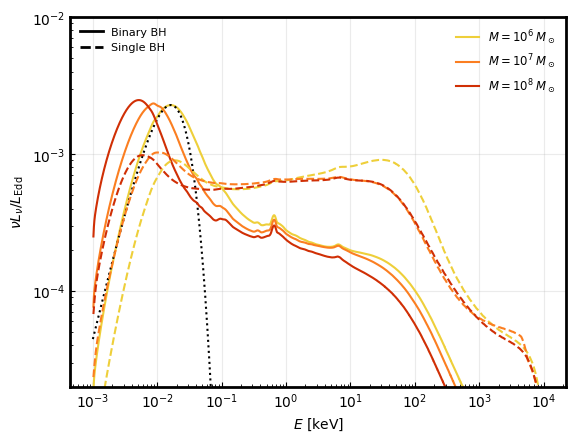}
    \caption{Solid angle-averaged broad band spectra as seen at infinity 
    for each of the black hole masses. The black dotted curve is a blackbody fit to the spectrum for $M = 10^6 M_\odot$. The dashed lines show the analogous spectra for single black holes out to $r=70\;r_g$ \citep{Nagele+2026a}. }
    \label{fig:spectra}
\end{figure*}

Fig.~\ref{fig:spectra} shows, for all three black hole masses, the angle-averaged spectra for both CBDs and single black holes (the latter taken from \citet{Nagele+2026a}); the accretion rate is the same $0.01 {\dot M}_{\rm Edd}$ in all six cases.  In addition, it shows a thermal spectrum fitted to the $10^6 M_\odot$ CBD spectrum.
For all three CBD spectra---but not the single black hole spectra--- most of the luminosity ($\simeq 60\%$) is in the UV thermal peak (Table \ref{tab:fits}),
and the remaining $\simeq 40\%$ is
radiated in a hard tail that extends to several tens of keV before rolling over. 
These fractions are all nearly the same because the luminosity is only slightly different from the volume integral of the simulation's cooling function, and the location of the photosphere is almost the same in all three.  The main reason why hard X-rays are much more prominent in the single black hole cases is that the CBD simulation excludes everything inside $15r_g$ and has very little matter between 15 and $40r_g$, while hard X-rays are largely made at the smaller radii.

There are two ways the spectrum depends on $M$, and both are weak dependences.  One
is to shift the thermal part of the spectrum to lower energies
as $M$ increases, 
scaling, as expected, $\propto M^{-1/4}$.
We caution that much of the thermal portion is radiated from disk regions where we have not performed explicit transfer solutions with \texttt{PTransX}, so the shape of the thermal portion is approximate.
The second is in the X-ray band, where the spectrum softens slightly with increasing $M$: the power-law index $\Gamma$ increases from 2.24 to 2.33 as $M$ grows from $10^6 M_\odot$ to $10^8 M_\odot$.
This effect is mostly due to the seed photons having a lower average energy in the higher mass cases.

In addition to the two continuum components, there are also two significant line features, OVIII~Ly$\alpha$ (at $\simeq 800$ eV) and Fe~K$\alpha$ (at $\simeq 6-7$~keV). The emergence of the latter from the optically-thick disk was discussed at length in the previous subsection.  Here we point out the equivalent widths for these lines in the observed spectra, which are modified from the disk surface spectrum by Compton scatters in the corona and relativistic shifts. We define these equivalent widths in terms of a power-law fit to the nearby continuum.  The energy ranges of the fit are
4 - 10~keV for Fe and 0.5 - 1~keV for O. With this definition, the Fe~K$\alpha$ lines have equivalent widths in the range $\sim 100-200$ eV, while 
OVIII~Ly$\alpha$ widths are $\sim30-50$ eV. These can be restated as fractions of the line energy in order to calibrate the lines' contrast to the continuum
(e.g., as in Fig.~\ref{fig:spectra_th}). The Fe~K$\alpha$ fractional width is $\sim 0.015-0.03$; for O~VIII it is $\sim 0.03-0.08$.
Note that these lines are produced in different places, with fluid frame Fe~K$\alpha$ luminosity peaking in the $40<r/r_g<50$ bin and OVIII~Ly$\alpha$ in the $60<r/r_g<80$ bin (Fig.~\ref{fig:flux_out}).  

\subsection{Angular dependence} 

\begin{figure}
\includegraphics[width=\linewidth]{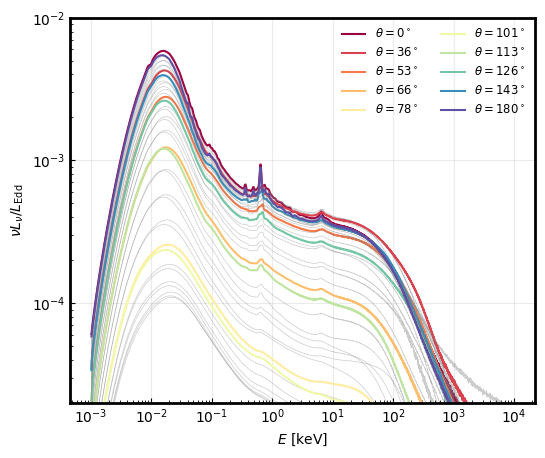}
    \caption{Observed spectra at different polar viewing angles for $M=10^6 M_\odot$. Grey lines show viewing angles intermediate to the colored curves. }
    \label{fig:spectra_th}
\end{figure}

Fig.~\ref{fig:spectra_th} shows the $\theta$ dependence of the spectrum for the $M=10^6\;\msun$ case. There are four main takeaways from this figure: a) the flux varies slowly with viewing angle for views well outside the equatorial plane, but near the plane there is very strong  self-obscuration due to the disk's vertical thickness; b) at any given moment, there can be a small difference in observed luminosity between the two hemispheres, c) the X-ray slope changes only slightly with inclination, and d) observed emission lines are much more prominent from 
directions near the polar axis
(as also found by \citet{Nagele+2026a,Nagele+2026b}).

One of the more interesting questions regarding CBDs is whether there are any periodic modulations of the observed spectra.
Although we analyze only a single snapshot of data, orbital rotation can make the spectrum seen by an observer at a fixed direction vary if the system's radiation pattern varies azimuthally.  The observed time-dependence is periodic for any spectral feature having two characteristics.  It must be made within a fairly narrow range of radii, so the entire emission region has the same orbital period; and it must change little over the duration of the observation.

\begin{figure}
\includegraphics[width=\linewidth]{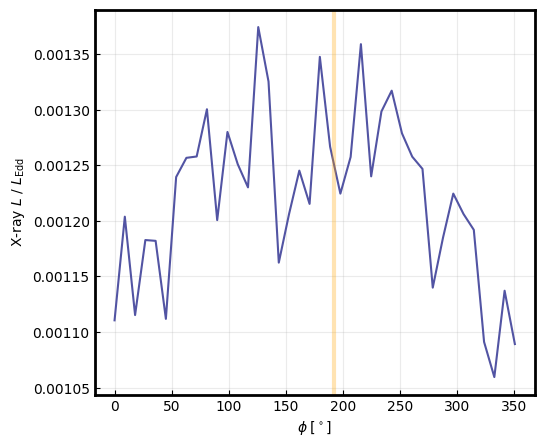}
    \caption{Integrated X-ray luminosity in Eddington units as a function of azimuthal viewing angle. The amber line shows observers directly opposite to the azimuth of the centroid of the disk heating rate (Sec. \ref{sec:discussion_implications}).}
    \label{fig:L_phi}
\end{figure}

The X-ray continuum may be such a feature.
In Fig. \ref{fig:L_phi} we show the integrated X-ray luminosity (photon energy above 0.5~keV) 
in Eddington units. The observed luminosity varies with azimuth, with a roughly 25\% difference between a peak and a trough separated by $\simeq \pi$~radians in phase.
This break in azimuthal symmetry is the result of two facts. The first is that the strongest sources of X-rays in the CBD are the spiral shocks riding just in front of the lump.
The second is that the disk's photosphere is tilted, so its surface normal, the direction of the radiated flux, has a radial component $-dH/dr \, \hat r$ (here $H/r$ is nearly constant, so this angle is nearly constant).
When photons are emitted from such a disk surface, the flux vector has a component in the $-\hat r$ direction.  As the lump and its attendant shocks orbit, the radially inward direction at their location rotates, and distant observers see the most X-rays when they are on the opposite side of the disk from the shocks.  In fact,
the azimuthal viewing angle of the peak in Fig.~\ref{fig:L_phi} is exactly in this direction.
This azimuthal variation of the X-ray luminosity is slight, and may be difficult to observed in practice, but it is derived from an effect which is both physical and persistent in time (Sec. \ref{sec:discussion_implications}).

\begin{figure}
\includegraphics[width=\linewidth]{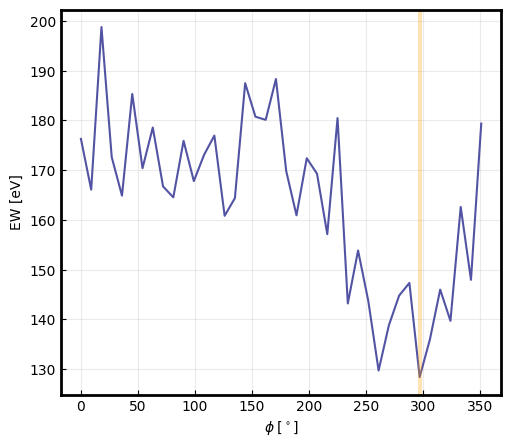}
    \caption{Fe K$\alpha$ equivalent width in eV as a function of azimuthal viewing angle. The amber line shows the azimuth of the centroid of the lump (Sec. \ref{sec:discussion_implications}).}
    \label{fig:FeKalpha_phi}
\end{figure}

Emission lines may be another periodic feature.
For Fe K$\alpha$, the main spectral change with azimuth is the level of the continuum, as discussed above, but the line profile also changes shape and location. To quantify this change, we plot the equivalent width of the observed line as a function of $\phi$ in Fig.~\ref{fig:FeKalpha_phi}.  
Over a range of phase $\simeq 4\pi/3$, the equivalent width is in the $160-190$ eV range, but drops by $\simeq 20\%$ for the remaining $\simeq 2\pi/3$ of phase.
This dip corresponds to the location of the lump, and we surmise that the large physical size of the lump (Fig. \ref{fig:azimuth_cf}) blocks 
some of the Fe~K$\alpha$ photons originating in the inner regions. 
Because this change in equivalent width is due to obscuration, it modulates not only K$\alpha$ emission from the CBD, but also K$\alpha$ from the minidisks.
We again caution that we have averaged the spectra over polar viewing angle in order to get the most robust possible statistics for the azimuthal variation. Studies with larger photon packet numbers 
should
be able to further decompose the azimuthal variation at multiple polar viewing angles.

\begin{figure}
\includegraphics[width=\linewidth]{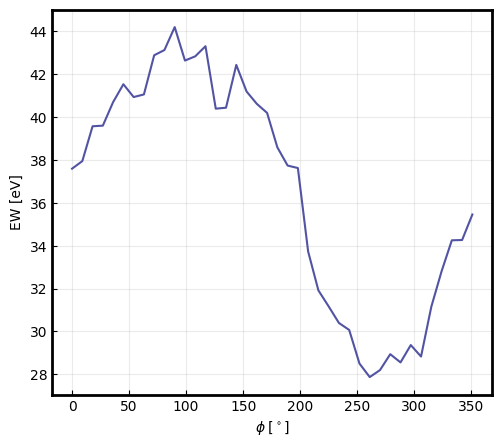}
    \caption{Same as Fig. \ref{fig:FeKalpha_phi} but for OVIII~Ly$\alpha$ instead of Fe~K$\alpha$.}
    \label{fig:O_phi}
\end{figure}

As shown in Fig.~\ref{fig:O_phi}, 
the angle-averaged equivalent width of OVIII~Ly$\alpha$ 
varies crudely sinusoidally with azimuthal angle.
The sinusoidal behavior can be understood as a combination of two effects we have already discussed. The first is that
the greatest OVIII~Ly$\alpha$ luminosity is concentrated in the region $r/r_g \simeq 70 \pm 10$.  For most of the azimuthal range of this region the photosphere has a wedge-like shape similar to the smaller radii, i.e., $dH/dr \sim const$.   However, at $r \gtrsim 70r_g$ and azimuthal angles $\phi \simeq 195^\circ \pm 45^\circ $, the photosphere is fairly flat, i.e., $H \simeq const.$
The wedge-like regions block photons radiated from the opposite side of the disk, but the flat-topped regions permit them to travel freely.
The result is to increase the line EW seen in the directions of the flat-topped regions.
The second 
is that the disk luminosity can be high in regions heated by spiral shocks. Earlier on, we focused on the shock around $50\;r_g$, but now we draw the reader's attention to $r=80\;r_g$, where there is enhanced dissipation at $\phi=300-330^\circ$, roughly opposite to the center of the flat-topped region.  This geometry creates a peak of equivalent width surrounding $\phi \sim 125^\circ$, as shown in Fig.~\ref{fig:O_phi}.
The Fe and O emission lines experience different geometrical effects because they are made at different radii. In particular, the flat photosphere at larger radii affects photons from
the OVIII Ly$\alpha$ emission region, but not Fe~K$\alpha$.

We caution, however, that 
the photospheric shape depends strongly on
the thermodynamics of the disk, which are treated with a rough approximation
in the \texttt{HARM3D} simulation. 
We cannot therefore make a strong prediction that regions with flat or tilted photospheres are generically present or are always placed where they are in this simulation; rather, we have demonstrated that once one views the disk as a 3D object, effects of this sort become possible.

\section{Discussion} \label{sec:discussion}

\begin{figure*}
 \includegraphics[width=\linewidth]{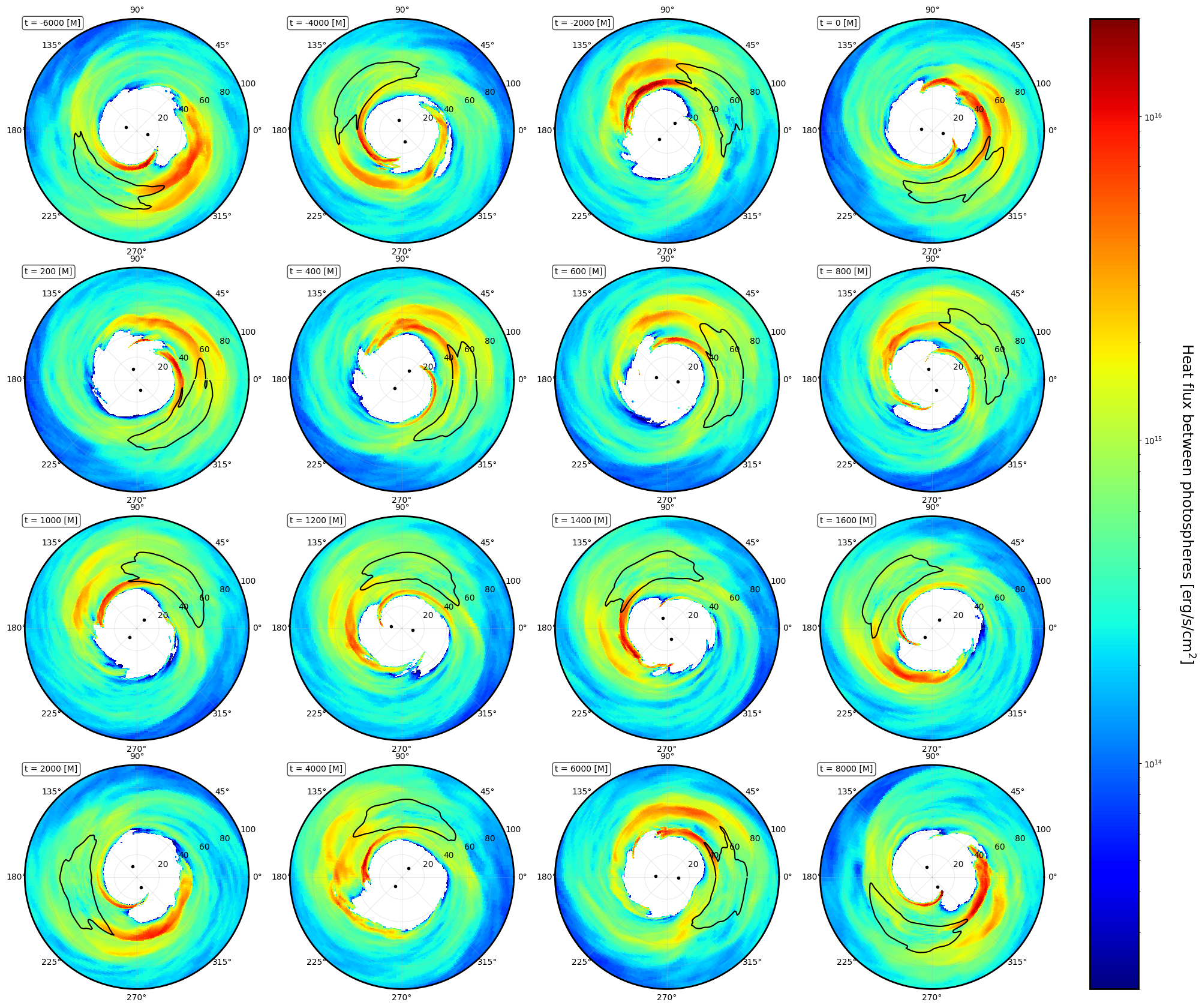}
    \caption{Heating rate between the top and bottom photospheres for 
    16
    \texttt{HARM3D} snapshots. White regions show locations where no
    photosphere exists. The black contours show regions of the disk with $\tau_s>30$; these regions are associated with the lump. The middle two rows are separated by 200 M and the upper and lower rows by 2000 M, compared to an orbital period of the lump of about 3000 M.}
    \label{fig:time_maps_fine}
\end{figure*}

\begin{figure}
 \includegraphics[width=\linewidth]{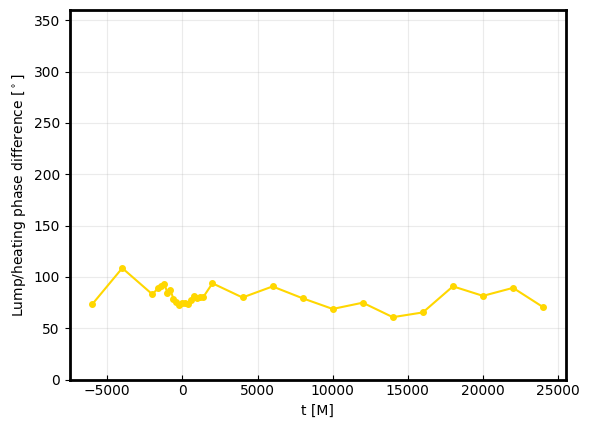}
    \caption{Relative phase between the lump and the disk heating which leads it. The points in this figure corresponds to the panels of Fig. \ref{fig:time_maps_fine}. }
    \label{fig:time_dphi}
\end{figure}

\subsection{Intrinsic
time 
dependence of the spectrum} \label{sec:discussion_implications}

In this paper, we have post-processed a single simulation snapshot and analyzed it in detail.  To indicate the degree to which CBD radiation may vary over time, we present, in
Fig.~\ref{fig:time_maps_fine}, maps of the integrated disk heating rate (cf. Fig.~\ref{fig:top_down}e).  These maps are each overlaid with a black contour showing $\tau_s=30$ to identify the lump (cf. Fig. \ref{fig:top_down}a).  The snapshots in the middle two rows are separated by $200\;M$ in time, covering $\sim 2/3$ of an orbit from $t=0$ to $t=2000M$. 
The times shown are relative to the time of the snapshot we used for post-processing. The images in the upper and lower rows are separated by $2000\;M$.

The snapshots
show that the disk heating consistently
takes place
in spiral filaments preceding the lump. These filaments are 
large amplitude spiral acoustic waves that have steepened into shocks (Sec.~\ref{sec:results_heating}).
These features trace the immediate post-shock gas in both the heating rate (Fig.~\ref{fig:time_maps_fine})
and gas pressure (Fig.~\ref{fig:curlB}).
They are found just ahead of the lump due to a timing coincidence between the torqued stream trajectories and the lump's orbit \citep{Shi2012ApJ...749..118S}.

In Fig. \ref{fig:time_dphi}, we plot the relative phase between the lump and the heated material. This is done by taking the centroid of the region of the disk with $\tau_s>30$ for the lump, and the centroid of the region of the disk with heating rate $>2\times10^{15}$ erg/s/cm$^2$, corresponding to the red/orange regions of Fig. \ref{fig:time_maps_fine}. The centroid of the heating rate leads the lump by about $90^\circ$. Using the centroids from the snapshot post-processed in this paper, we overlaid the lump position in Fig. \ref{fig:FeKalpha_phi} and $180^\circ$ minus the heating position in Fig. \ref{fig:L_phi}. As expected, the lump coincides with the dip in the Fe K$\alpha$ equivalent width, and the maximum heating location is at an azimuth opposite to the direction of the peak X-ray luminosity.

We thus suggest that X-ray observations of active galactic nuclei should be used to search for 
modulations of both the X-ray continuum luminosity and the Fe~K$\alpha$ equivalent width with the same period, but with the K$\alpha$ minimum appearing at a phase $\approx \pi/2$ before the continuum maximum.  The relevant scale for the observed period would be the period of the lump $\sim 0.5 (M/10^8 M_\odot) (a/100r_g)^{3/2}$~yr (note that the binary lifetime scales $\propto (a/M)^4 M$). 
If such a system were found, it would be a strong candidate for localized gravitational wave follow-up via PTA \citep[e.g.][]{Agarwal2026ApJ...998L..11A}. Due to the small amplitude of the modulation in both the continuum and the line, as well as the sensitivity level of the PTA method, this procedure would be viable only for relatively nearby systems.

\subsection{Comparison to previous work} \label{sec:discussion_comparison}

Previous studies of the spectrum originating from SMBH binary systems have focused on three main areas: a) the thermal continuum, b) the X-ray continuum, and c) X-ray emission lines. 

The thermal continuum is sometimes thought to have a notch in it, which would be evidence for the existence of a binary \citep{Roedig2014ApJ...785..115R,Farris2015MNRAS.446L..36F,Tang2018MNRAS.476.2249T}.
The notch is created when matter crosses the cavity from the CBD to the minidisks, but does not radiate thermally the potential energy it loses.
For this notch to be noticeable requires that the outer edges of the minidisks and the CBD contribute comparable thermal luminosity \citep{dAscoli2018ApJ...865..140D}, but well separated in temperature.

\citet{Farris2015MNRAS.446L..36F} and \citet{Tang2018MNRAS.476.2249T} post-processed 2D viscous hydrodynamics simulations of SMBH binaries (including both the CBD and the minidisks), assuming purely thermal emission, and both found a notch in their predicted spectra. \citet{Tiwari2025ApJ...986..158T} ran radiation MHD simulations of a CBD with accretion streams in 3D but no minidisks, similar to our simulations, and post-processed their simulation assuming thermal spectra. Their spectra peaked at $\simeq 4$~eV.  Using classical disk theory to scale this result from their central mass, accretion rate, and separation  ($2 \times 10^7 M_\odot$, $0.15 {\dot M}_{\rm Edd}$, and $100 r_g)$ to ours predicts that their peak would be at an energy $\simeq 1/2$ ours, which is almost exactly the actual ratio.  Also like ours,
their predicted spectrum falls steeply at higher energies. However, because they assumed purely thermal radiation, they were unable to make a statement about hard X-ray emission.
\citet{dAscoli2018ApJ...865..140D} and \citet{Gutierrez2022ApJ...928..137G} used the \texttt{BOTHROS} code to postprocess \texttt{HARM3D} simulations (such as the one we post-process), assuming that the optically thick material emitted purely thermally, and that the optically thin material radiated with a typical AGN X-ray spectrum. They found no notch because the minidisks in a binary of non-spinning black holes separated by only $20M$ radiate somewhat less luminosity than the CBD.

We do not include the minidisks, so we would not expect to see a notch in our spectra, but we do see that $\nu F_\nu$ drops by just short of an order of magnitude in the energy decade to the right of the peak (Fig.~\ref{fig:spectra}), indicating that a notch could be present were we to include the minidisks. The thermal parts of our spectra are not directly comparable to those of \citet{Tang2018MNRAS.476.2249T} because their accretion rate is much higher than ours, but are qualitatively consistent with the results of \citet{dAscoli2018ApJ...865..140D,Gutierrez2022ApJ...928..137G,Tiwari2025ApJ...986..158T}.

For non-thermal emission, we find that our X-ray luminosity is significantly higher relative to the thermal luminosity than found by previous works \citep{dAscoli2018ApJ...865..140D,Gutierrez2022ApJ...928..137G}. This difference may be partly due 
to a lower accretion rate in our work, but the major difference is that we do not assume purely thermal emission from within the disk.  Instead we predict the disk spectrum using frequency-dependent opacities and emissivities derived from atomic physics; the resulting spectra become purely thermal only in the limit of very high density and optical depth.  As a result,
some regions of the disk actually emit spectra with hard power laws (e.g. Fig. \ref{fig:flux_out_comparison}).

Actual observed spectra would include the minidisks.  Although we have not computed their spectra in detail, we can make rough estimates of the relative importance of the minidisks' radiation.
In bolometric terms,
the observed luminosity of the snapshot we analyze
($L/L_{\rm Edd}=0.007$)
is comparable to the luminosity
produced by the innermost $70r_g$ of an  accretion disk around a single black hole with the same nominal accretion rate, $\dot m = 0.01$,
which  varies within the range $\in [0.005,0.02]$ \citet{Liu+2025}. 
In a binary of the separation we treat, $20r_g$, the minidisks' luminosity would be rather less than this single black hole estimate because they
extend out to only $r=12\;r_g$ (for the individual black hole $r_g = 0.5M$).
Thus, in the context of the thermal portion, notch formation may be marginal in such systems, but this conclusion is sensitive to both the separation of the binary and the spins of the black holes.  The importance of the minidisks to the non-thermal portion depends strongly on what fraction of their bolometric light goes into the X-ray band.  If it is the usual $\sim 10\%$, the contributions of the minidisks and the CBD would, for our choices of separation and black hole spins, would be comparable.  On the other hand, if the minidisks radiate most of their power in the X-ray band, it would dominate even the relatively hard spectrum we find is radiated by the CBD.

Finally, recent work has suggested that SMBH binaries could be identified through Fe~K$\alpha$ line profiles from the minidisks \citep{Malewicz2025ApJ...989..190M,Malewicz2026arXiv260804961M}. They considered binaries at a separation of $a=100\;r_g$ and used X-ray reflection modeling to predict the iron line coming from a minidisk, after which they applied an orbital Doppler shift to the component of the line emerging from each minidisk. In this paper, we identify a  periodic modulation of the iron line (and the oxygen line) due to the orbital motion of the inner edge of the cavity, but this modulation has its origin in disk geometry, not the Doppler shift.  This modulation has a frequency $0.1-0.2\times$ the binary frequency. If a Fourier analysis of a binary candidate revealed periodicity at two frequencies separated by a factor of $0.1-0.2$, it would be an indicator
of a binary. If, on the other hand, it revealed periodic variation at only one frequency, it would raise the question of whether that frequency should be associated with the minidisks or the CBD.

\begin{table}
\centering
\begin{tblr}{
colspec={ccccc}
}  Log $M/\msun$ & $L/L_{\rm Edd}$ & $L_{\rm non\;thermal} / L$ & Log $T_{\rm obs}$ & $\Gamma$ \\ \hline
6 & 0.0067 & 0.43 & 4.83 & 2.24 \\
7 & 0.0067 & 0.42 & 4.58 & 2.27 \\
8 & 0.0067 & 0.41 & 4.36 & 2.33 \\
\hline
\end{tblr}
\caption{Summary of observed spectra for each black hole mass. Observed luminosity, non-thermal fraction, blackbody temperature fit to the observed angle-averaged spectrum, and photon index of the observed angle-averaged spectrum. $L/L_{\rm Edd}$ is constant for all three masses, as each makes use of the same underlying \texttt{HARM3D} snapshot. The non thermal fraction, observed blackbody temperature, and photon index vary slightly with mass. 
} \label{tab:fits}
\end{table}

\section{Conclusions} \label{sec:conclusion}

We have conducted the first radiative transfer postprocessing with detailed radiation physics of GRMHD simulations of a circumbinary disk surrounding a supermassive black hole binary. We solved for the radiation field under the assumption of thermal balance
throughout the system
and additionally under the assumption of ionization balance within the optically thick diskbody.  Because our method joins a dynamic GR metric and MHD to 3D radiation transfer and ionization balance, we can probe the differences in radiation spectra generated at different locations throughout the disk.

As is demonstrated by the variety of 
local spectra
seen in Fig.~\ref{fig:flux_out_comparison}, this power enabled the identification of multiple distinct regions spanning a wide range of physical conditions.  These regions often exhibit complex geometry and asymmetry, as well as strongly contrasting thermal properties. 
Some, like the lump, maintain a temperature $\sim 10^5 - 10^6$~K, as follows from the commonly-adopted assumption of LTE for accretion disks this close to a supermassive black hole.  However, others  have gas temperatures $\sim 10^8$~K despite being nearly as optically thick.  They are able to do so because the absorption optical depth is much less than the scattering optical depth. The heating mechanism for regions like this in the accretion streams is magnetic dissipation, but a similar thermodynamic state is reached in spiral shocks internal to the disk. 
The end result is that a distant observer viewing only the circumbinary disk would see not only a thermal spectrum, but also a strong X-ray tail.

Our approach, which rests on physical processes---3D density maps from 3D GRMHD, detailed atomic physics, multi-zone and multi-angle radiation transfer---also yielded several other potentially observational diagnostics, including a number of periodic modulations with periods on the scale of the binary orbital period.
When the CBD's inner edge has a sizable density concentration (the ``lump"),
much of the X-ray emission comes from a region lying immediately ahead of the lump in terms of orbital motion.  This break in axisymmetry, combined with the tilted surface of the disk photosphere, can cause a periodic modulation in the X-ray continuum. 
In addition, the Fe~K$\alpha$ equivalent width dips when the lump passes between the central part of the disk and the observer (Fig. \ref{fig:FeKalpha_phi}). 
Another periodicity is observed in the equivalent width of O~Ly$\alpha$ (Fig. \ref{fig:O_phi}).
These are the {\it only} periodic signals in circumbinary disks to date that
emerged from self-consistent spectral 
postprocessing of GRMHD simulations of circumbinary disks.

\begin{center}  \label{Sec:ack}
    \textbf{Acknowledgments}
\end{center}

This work was partially supported by NASA TCAN grant 80NSSC24K0100.

\bibliography{references}{}

\begin{thebibliography}{}
\expandafter\ifx\csname natexlab\endcsname\relax\def\natexlab#1{#1}\fi
\providecommand{\url}[1]{\href{#1}{#1}}
\providecommand{\dodoi}[1]{doi:~\href{http://doi.org/#1}{\nolinkurl{#1}}}
\providecommand{\doeprint}[1]{\href{http://ascl.net/#1}{\nolinkurl{http://ascl.net/#1}}}
\providecommand{\doarXiv}[1]{\href{https://arxiv.org/abs/#1}{\nolinkurl{https://arxiv.org/abs/#1}}}

\bibitem[{{Abbott} {et~al.}(2017{\natexlab{a}}){Abbott}, {Abbott}, {Abbott}, {Acernese}, {Ackley}, {Adams}, {Adams}, {Addesso}, {Adhikari}, {Adya}, {Affeldt}, {Afrough}, {Agarwal}, {Agathos}, {Agatsuma}, {Aggarwal}, {Aguiar}, {Aiello}, {Ain}, {Ajith}, {Allen}, {Allen}, {Allocca}, {Altin}, {Amato}, {Ananyeva}, {Anderson}, {Anderson}, {Angelova}, {Antier}, {Appert}, {Arai}, {Araya}, {Areeda}, {Arnaud}, {Arun}, {Ascenzi}, {Ashton}, {Ast}, {Aston}, {Astone}, {Atallah}, {Aufmuth}, {Aulbert}, {AultONeal}, {Austin}, {Avila-Alvarez}, {Babak}, {Bacon}, {Bader}, {Bae}, {Baker}, {Baldaccini}, {Ballardin}, {Ballmer}, {Banagiri}, {Barayoga}, {Barclay}, {Barish}, {Barker}, {Barkett}, {Barone}, {Barr}, {Barsotti}, {Barsuglia}, {Barta}, {Barthelmy}, {Bartlett}, {Bartos}, {Bassiri}, {Basti}, {Batch}, {Bawaj}, {Bayley}, {Bazzan}, {B{\'e}csy}, {Beer}, {Bejger}, {Belahcene}, {Bell}, {Berger}, {Bergmann}, {Bero}, {Berry}, {Bersanetti}, {Bertolini}, {Betzwieser}, {Bhagwat}, {Bhandare}, {Bilenko}, {Billingsley}, {Billman}, {Birch},
  {Birney}, {Birnholtz}, {Biscans}, {Biscoveanu}, {Bisht}, {Bitossi}, {Biwer}, {Bizouard}, {Blackburn}, {Blackman}, {Blair}, {Blair}, {Blair}, {Bloemen}, {Bock}, {Bode}, {Boer}, {Bogaert}, {Bohe}, {Bondu}, {Bonilla}, {Bonnand}, {Boom}, {Bork}, {Boschi}, {Bose}, {Bossie}, {Bouffanais}, {Bozzi}, {Bradaschia}, {Brady}, {Branchesi}, {Brau}, {Briant}, {Brillet}, {Brinkmann}, {Brisson}, {Brockill}, {Broida}, {Brooks}, {Brown}, {Brown}, {Brunett}, {Buchanan}, {Buikema}, {Bulik}, {Bulten}, {Buonanno}, {Buskulic}, {Buy}, {Byer}, {Cabero}, {Cadonati}, {Cagnoli}, {Cahillane}, {Calder{\'o}n Bustillo}, {Callister}, {Calloni}, {Camp}, {Canepa}, {Canizares}, {Cannon}, {Cao}, {Cao}, {Capano}, {Capocasa}, {Carbognani}, {Caride}, {Carney}, {Casanueva Diaz}, {Casentini}, {Caudill}, {Cavagli{\`a}}, {Cavalier}, {Cavalieri}, {Cella}, {Cepeda}, {Cerd{\'a}-Dur{\'a}n}, {Cerretani}, {Cesarini}, {Chamberlin}, {Chan}, {Chao}, {Charlton}, {Chase}, {Chassande-Mottin}, {Chatterjee}, {Chatziioannou}, {Cheeseboro}, {Chen}, {Chen}, {Chen},
  {Cheng}, {Chia}, {Chincarini}, {Chiummo}, {Chmiel}, {Cho}, {Cho}, {Chow}, {Christensen}, {Chu}, {Chua}, {Chua}, {Chung}, {Chung}, \& {Ciani}}]{Abbott2017ApJ...848L..12A}
{Abbott}, B.~P., {Abbott}, R., {Abbott}, T.~D., {et~al.} 2017{\natexlab{a}}, \apjl, 848, L12, \dodoi{10.3847/2041-8213/aa91c9}

\bibitem[{{Abbott} {et~al.}(2017{\natexlab{b}}){Abbott}, {Abbott}, {Abbott}, {Acernese}, {Ackley}, {Adams}, {Adams}, {Addesso}, {Adhikari}, {Adya}, {Affeldt}, {Afrough}, {Agarwal}, {Agathos}, {Agatsuma}, {Aggarwal}, {Aguiar}, {Aiello}, {Ain}, {Ajith}, {Allen}, {Allen}, {Allocca}, {Altin}, {Amato}, {Ananyeva}, {Anderson}, {Anderson}, {Angelova}, {Antier}, {Appert}, {Arai}, {Araya}, {Areeda}, {Arnaud}, {Arun}, {Ascenzi}, {Ashton}, {Ast}, {Aston}, {Astone}, {Atallah}, {Aufmuth}, {Aulbert}, {Aultoneal}, {Austin}, {Avila-Alvarez}, {Babak}, {Bacon}, {Bader}, {Bae}, {Baker}, {Baldaccini}, {Ballardin}, {Ballmer}, {Banagiri}, {Barayoga}, {Barclay}, {Barish}, {Barker}, {Barkett}, {Barone}, {Barr}, {Barsotti}, {Barsuglia}, {Barta}, {Bartlett}, {Bartos}, {Bassiri}, {Basti}, {Batch}, {Bawaj}, {Bayley}, {Bazzan}, {B{\'e}csy}, {Beer}, {Bejger}, {Belahcene}, {Bell}, {Berger}, {Bergmann}, {Bero}, {Berry}, {Bersanetti}, {Bertolini}, {Betzwieser}, {Bhagwat}, {Bhandare}, {Bilenko}, {Billingsley}, {Billman}, {Birch}, {Birney},
  {Birnholtz}, {Biscans}, {Biscoveanu}, {Bisht}, {Bitossi}, {Biwer}, {Bizouard}, {Blackburn}, {Blackman}, {Blair}, {Blair}, {Blair}, {Bloemen}, {Bock}, {Bode}, {Boer}, {Bogaert}, {Bohe}, {Bondu}, {Bonilla}, {Bonnand}, {Boom}, {Bork}, {Boschi}, {Bose}, {Bossie}, {Bouffanais}, {Bozzi}, {Bradaschia}, {Brady}, {Branchesi}, {Brau}, {Briant}, {Brillet}, {Brinkmann}, {Brisson}, {Brockill}, {Broida}, {Brooks}, {Brown}, {Brown}, {Brunett}, {Buchanan}, {Buikema}, {Bulik}, {Bulten}, {Buonanno}, {Buskulic}, {Buy}, {Byer}, {Cabero}, {Cadonati}, {Cagnoli}, {Cahillane}, {Bustillo}, {Callister}, {Calloni}, {Camp}, {Canepa}, {Canizares}, {Cannon}, {Cao}, {Cao}, {Capano}, {Capocasa}, {Carbognani}, {Caride}, {Carney}, {Diaz}, {Casentini}, {Caudill}, {Cavagli{\`a}}, {Cavalier}, {Cavalieri}, {Cella}, {Cepeda}, {Cerd{\'a}-Dur{\'a}n}, {Cerretani}, {Cesarini}, {Chamberlin}, {Chan}, {Chao}, {Charlton}, {Chase}, {Chassande-Mottin}, {Chatterjee}, {Chatziioannou}, {Cheeseboro}, {Chen}, {Chen}, {Chen}, {Cheng}, {Chia}, {Chincarini},
  {Chiummo}, {Chmiel}, {Cho}, {Cho}, {Chow}, {Christensen}, {Chu}, {Chua}, {Chua}, {Chung}, {Chung}, {Ciani}, \& {Ciolfi}}]{Abbott2017Natur.551...85A}
---. 2017{\natexlab{b}}, \nat, 551, 85, \dodoi{10.1038/nature24471}

\bibitem[{{Agarwal} {et~al.}(2026){Agarwal}, {Agazie}, {Anumarlapudi}, {Archibald}, {Arzoumanian}, {Baier}, {Baker}, {B{\'e}csy}, {Blecha}, {Brazier}, {Brook}, {Burke-Spolaor}, {Burnette}, {Case}, {Casey-Clyde}, {Chang}, {Charisi}, {Chatterjee}, {Cohen}, {Coppi}, {Cordes}, {Cornish}, {Crawford}, {Cromartie}, {Crowter}, {Decesar}, {Demorest}, {Deng}, {Dey}, {Dolch}, {D'Orazio}, {Eisenberg}, {Ferrara}, {Doskoch}, {Fiore}, {Fonseca}, {Freedman}, {Gardiner}, {Garver-Daniels}, {Gentile}, {Gersbach}, {Glaser}, {Graham}, {Good}, {G{\"u}ltekin}, {Harris}, {Hazboun}, {Hutchison}, {Jennings}, {Johnson}, {Jones}, {Kaplan}, {Kelley}, {Kerr}, {Key}, {Laal}, {Lam}, {Lamb}, {Larsen}, {Lazio}, {Lewandowska}, {Liu}, {Lorimer}, {Luo}, {Lynch}, {Ma}, {Madison}, {Matt}, {McEwen}, {McKee}, {McLaughlin}, {McMann}, {Meyers}, {Meyers}, {Mingarelli}, {Mitridate}, {Natarajan}, {Ng}, {Nice}, {Nichols}, {Ocker}, {Olum}, {Pennucci}, {Perera}, {Petrov}, {Pol}, {Radovan}, {Ransom}, {Ray}, {Romano}, {Runnoe}, {Saffer}, {Sardesai},
  {Schmiedekamp}, {Schmiedekamp}, {Schmitz}, {Semenzato}, {Shapiro-Albert}, {Shivakumar}, {Siemens}, {Simon}, {Sosa Fiscella}, {Stairs}, {Stinebring}, {Stovall}, {Susobhanan}, {Swiggum}, {Taylor}, {Taylor}, {Thompson}, {Turner}, {Vallisneri}, {van Haasteren}, {Vigeland}, {Wahl}, {Willson}, {Wilson}, {Witt}, {Wright}, {Young}, {Zheng}, \& {Nanograv Collaboration}}]{Agarwal2026ApJ...998L..11A}
{Agarwal}, N., {Agazie}, G., {Anumarlapudi}, A., {et~al.} 2026, \apjl, 998, L11, \dodoi{10.3847/2041-8213/ae3719}

\bibitem[{{Agazie} {et~al.}(2023){Agazie}, {Anumarlapudi}, {Archibald}, {Baker}, {B{\'e}csy}, {Blecha}, {Bonilla}, {Brazier}, {Brook}, {Burke-Spolaor}, {Burnette}, {Case}, {Casey-Clyde}, {Charisi}, {Chatterjee}, {Chatziioannou}, {Cheeseboro}, {Chen}, {Cohen}, {Cordes}, {Cornish}, {Crawford}, {Cromartie}, {Crowter}, {Cutler}, {D'Orazio}, {Decesar}, {Degan}, {Demorest}, {Deng}, {Dolch}, {Drachler}, {Ferrara}, {Fiore}, {Fonseca}, {Freedman}, {Gardiner}, {Garver-Daniels}, {Gentile}, {Gersbach}, {Glaser}, {Good}, {G{\"u}ltekin}, {Hazboun}, {Hourihane}, {Islo}, {Jennings}, {Johnson}, {Jones}, {Kaiser}, {Kaplan}, {Kelley}, {Kerr}, {Key}, {Laal}, {Lam}, {Lamb}, {Lazio}, {Lewandowska}, {Littenberg}, {Liu}, {Luo}, {Lynch}, {Ma}, {Madison}, {McEwen}, {McKee}, {McLaughlin}, {McMann}, {Meyers}, {Meyers}, {Mingarelli}, {Mitridate}, {Natarajan}, {Ng}, {Nice}, {Ocker}, {Olum}, {Pennucci}, {Perera}, {Petrov}, {Pol}, {Radovan}, {Ransom}, {Ray}, {Romano}, {Runnoe}, {Sardesai}, {Schmiedekamp}, {Schmiedekamp}, {Schmitz},
  {Schult}, {Shapiro-Albert}, {Siemens}, {Simon}, {Siwek}, {Stairs}, {Stinebring}, {Stovall}, {Sun}, {Susobhanan}, {Swiggum}, {Taylor}, {Taylor}, {Turner}, {Unal}, {Vallisneri}, {Vigeland}, {Wachter}, {Wahl}, {Wang}, {Witt}, {Wright}, {Young}, \& {Nanograv Collaboration}}]{Agazie2023ApJ...952L..37A}
{Agazie}, G., {Anumarlapudi}, A., {Archibald}, A.~M., {et~al.} 2023, \apjl, 952, L37, \dodoi{10.3847/2041-8213/ace18b}

\bibitem[{{Amaro-Seoane} {et~al.}(2023){Amaro-Seoane}, {Andrews}, {Arca Sedda}, {Askar}, {Baghi}, {Balasov}, {Bartos}, {Bavera}, {Bellovary}, {Berry}, {Berti}, {Bianchi}, {Blecha}, {Blondin}, {Bogdanovi{\'c}}, {Boissier}, {Bonetti}, {Bonoli}, {Bortolas}, {Breivik}, {Capelo}, {Caramete}, {Cattorini}, {Charisi}, {Chaty}, {Chen}, {Chru{\'s}li{\'n}ska}, {Chua}, {Church}, {Colpi}, {D'Orazio}, {Danielski}, {Davies}, {Dayal}, {De Rosa}, {Derdzinski}, {Destounis}, {Dotti}, {Du{\c{t}}an}, {Dvorkin}, {Fabj}, {Foglizzo}, {Ford}, {Fouvry}, {Franchini}, {Fragos}, {Fryer}, {Gaspari}, {Gerosa}, {Graziani}, {Groot}, {Habouzit}, {Haggard}, {Haiman}, {Han}, {Istrate}, {Johansson}, {Khan}, {Kimpson}, {Kokkotas}, {Kong}, {Korol}, {Kremer}, {Kupfer}, {Lamberts}, {Larson}, {Lau}, {Liu}, {Lloyd-Ronning}, {Lodato}, {Lupi}, {Ma}, {Maccarone}, {Mandel}, {Mangiagli}, {Mapelli}, {Mathis}, {Mayer}, {McGee}, {McKernan}, {Miller}, {Mota}, {Mumpower}, {Nasim}, {Nelemans}, {Noble}, {Pacucci}, {Panessa}, {Paschalidis}, {Pfister}, {Porquet},
  {Quenby}, {Ricarte}, {R{\"o}pke}, {Regan}, {Rosswog}, {Ruiter}, {Ruiz}, {Runnoe}, {Schneider}, {Schnittman}, {Secunda}, {Sesana}, {Seto}, {Shao}, {Shapiro}, {Sopuerta}, {Stone}, {Suvorov}, {Tamanini}, {Tamfal}, {Tauris}, {Temmink}, {Tomsick}, {Toonen}, {Torres-Orjuela}, {Toscani}, {Tsokaros}, {Unal}, {V{\'a}zquez-Aceves}, {Valiante}, {van Putten}, {van Roestel}, {Vignali}, {Volonteri}, {Wu}, {Younsi}, {Yu}, {Zane}, {Zwick}, {Antonini}, {Baibhav}, {Barausse}, {Bonilla Rivera}, {Branchesi}, {Branduardi-Raymont}, {Burdge}, {Chakraborty}, {Cuadra}, {Dage}, {Davis}, {de Mink}, {Decarli}, {Doneva}, {Escoffier}, {Gandhi}, {Haardt}, {Lousto}, {Nissanke}, {Nordhaus}, {O'Shaughnessy}, {Portegies Zwart}, {Pound}, {Schussler}, {Sergijenko}, {Spallicci}, {Vernieri}, \& {Vigna-G{\'o}mez}}]{Amaro-Seoane2023LRR....26....2A}
{Amaro-Seoane}, P., {Andrews}, J., {Arca Sedda}, M., {et~al.} 2023, Living Reviews in Relativity, 26, 2, \dodoi{10.1007/s41114-022-00041-y}

\bibitem[{{Arnett} {et~al.}(1989){Arnett}, {Bahcall}, {Kirshner}, \& {Woosley}}]{Arnett1989ARA&A..27..629A}
{Arnett}, W.~D., {Bahcall}, J.~N., {Kirshner}, R.~P., \& {Woosley}, S.~E. 1989, \araa, 27, 629, \dodoi{10.1146/annurev.aa.27.090189.003213}

\bibitem[{{Artymowicz} \& {Lubow}(1996)}]{Artymowicz1996ApJ...467L..77A}
{Artymowicz}, P., \& {Lubow}, S.~H. 1996, \apjl, 467, L77, \dodoi{10.1086/310200}

\bibitem[{{Bogdanovi{\'c}} {et~al.}(2022){Bogdanovi{\'c}}, {Miller}, \& {Blecha}}]{Bogdanovic2022LRR....25....3B}
{Bogdanovi{\'c}}, T., {Miller}, M.~C., \& {Blecha}, L. 2022, Living Reviews in Relativity, 25, 3, \dodoi{10.1007/s41114-022-00037-8}

\bibitem[{{Bowen} {et~al.}(2017){Bowen}, {Campanelli}, {Krolik}, {Mewes}, \& {Noble}}]{Bowen2017ApJ...838...42B}
{Bowen}, D.~B., {Campanelli}, M., {Krolik}, J.~H., {Mewes}, V., \& {Noble}, S.~C. 2017, \apj, 838, 42, \dodoi{10.3847/1538-4357/aa63f3}

\bibitem[{{Bowen} {et~al.}(2019){Bowen}, {Mewes}, {Noble}, {Avara}, {Campanelli}, \& {Krolik}}]{Bowen2019ApJ...879...76B}
{Bowen}, D.~B., {Mewes}, V., {Noble}, S.~C., {et~al.} 2019, \apj, 879, 76, \dodoi{10.3847/1538-4357/ab2453}

\bibitem[{{Bright} \& {Paschalidis}(2023)}]{Bright2023MNRAS.520..392B}
{Bright}, J.~C., \& {Paschalidis}, V. 2023, \mnras, 520, 392, \dodoi{10.1093/mnras/stad091}

\bibitem[{{Charisi} {et~al.}(2016){Charisi}, {Bartos}, {Haiman}, {Price-Whelan}, {Graham}, {Bellm}, {Laher}, \& {M{\'a}rka}}]{Charisi2016MNRAS.463.2145C}
{Charisi}, M., {Bartos}, I., {Haiman}, Z., {et~al.} 2016, \mnras, 463, 2145, \dodoi{10.1093/mnras/stw1838}

\bibitem[{{Chen} {et~al.}(2020){Chen}, {Liu}, {Liao}, {Holgado}, {Guo}, {Gruendl}, {Morganson}, {Shen}, {Zhang}, {Abbott}, {Aguena}, {Allam}, {Avila}, {Bertin}, {Bhargava}, {Brooks}, {Burke}, {Carnero Rosell}, {Carollo}, {Carrasco Kind}, {Carretero}, {Costanzi}, {da Costa}, {Davis}, {De Vicente}, {Desai}, {Diehl}, {Doel}, {Everett}, {Flaugher}, {Friedel}, {Frieman}, {Garc{\'\i}a-Bellido}, {Gaztanaga}, {Glazebrook}, {Gruen}, {Gutierrez}, {Hinton}, {Hollowood}, {James}, {Kim}, {Kuehn}, {Kuropatkin}, {Lewis}, {Lidman}, {Lima}, {Maia}, {March}, {Marshall}, {Menanteau}, {Miquel}, {Palmese}, {Paz-Chinch{\'o}n}, {Plazas}, {Sanchez}, {Schubnell}, {Serrano}, {Sevilla-Noarbe}, {Smith}, {Suchyta}, {Swanson}, {Tarle}, {Tucker}, {Norbert Varga}, \& {Walker}}]{Chen2020MNRAS.499.2245C}
{Chen}, Y.-C., {Liu}, X., {Liao}, W.-T., {et~al.} 2020, \mnras, 499, 2245, \dodoi{10.1093/mnras/staa2957}

\bibitem[{{Chen} {et~al.}(2024){Chen}, {Zhai}, {Liu}, {Guo}, {Peng}, {Li}, {Songsheng}, {Du}, {Hu}, \& {Wang}}]{Chen2024MNRAS.52712154C}
{Chen}, Y.-J., {Zhai}, S., {Liu}, J.-R., {et~al.} 2024, \mnras, 527, 12154, \dodoi{10.1093/mnras/stad3981}

\bibitem[{{Cowperthwaite} {et~al.}(2017){Cowperthwaite}, {Berger}, {Villar}, {Metzger}, {Nicholl}, {Chornock}, {Blanchard}, {Fong}, {Margutti}, {Soares-Santos}, {Alexander}, {Allam}, {Annis}, {Brout}, {Brown}, {Butler}, {Chen}, {Diehl}, {Doctor}, {Drout}, {Eftekhari}, {Farr}, {Finley}, {Foley}, {Frieman}, {Fryer}, {Garc{\'\i}a-Bellido}, {Gill}, {Guillochon}, {Herner}, {Holz}, {Kasen}, {Kessler}, {Marriner}, {Matheson}, {Neilsen}, {Quataert}, {Palmese}, {Rest}, {Sako}, {Scolnic}, {Smith}, {Tucker}, {Williams}, {Balbinot}, {Carlin}, {Cook}, {Durret}, {Li}, {Lopes}, {Louren{\c{c}}o}, {Marshall}, {Medina}, {Muir}, {Mu{\~n}oz}, {Sauseda}, {Schlegel}, {Secco}, {Vivas}, {Wester}, {Zenteno}, {Zhang}, {Abbott}, {Banerji}, {Bechtol}, {Benoit-L{\'e}vy}, {Bertin}, {Buckley-Geer}, {Burke}, {Capozzi}, {Carnero Rosell}, {Carrasco Kind}, {Castander}, {Crocce}, {Cunha}, {D'Andrea}, {da Costa}, {Davis}, {DePoy}, {Desai}, {Dietrich}, {Drlica-Wagner}, {Eifler}, {Evrard}, {Fernandez}, {Flaugher}, {Fosalba}, {Gaztanaga}, {Gerdes},
  {Giannantonio}, {Goldstein}, {Gruen}, {Gruendl}, {Gutierrez}, {Honscheid}, {Jain}, {James}, {Jeltema}, {Johnson}, {Johnson}, {Kent}, {Krause}, {Kron}, {Kuehn}, {Nuropatkin}, {Lahav}, {Lima}, {Lin}, {Maia}, {March}, {Martini}, {McMahon}, {Menanteau}, {Miller}, {Miquel}, {Mohr}, {Neilsen}, {Nichol}, {Ogando}, {Plazas}, {Roe}, {Romer}, {Roodman}, {Rykoff}, {Sanchez}, {Scarpine}, {Schindler}, {Schubnell}, {Sevilla-Noarbe}, {Smith}, {Smith}, {Sobreira}, {Suchyta}, {Swanson}, {Tarle}, {Thomas}, {Thomas}, {Troxel}, {Vikram}, {Walker}, {Wechsler}, {Weller}, {Yanny}, \& {Zuntz}}]{Cowperthwaite2017ApJ...848L..17C}
{Cowperthwaite}, P.~S., {Berger}, E., {Villar}, V.~A., {et~al.} 2017, \apjl, 848, L17, \dodoi{10.3847/2041-8213/aa8fc7}

\bibitem[{{d'Ascoli} {et~al.}(2018){d'Ascoli}, {Noble}, {Bowen}, {Campanelli}, {Krolik}, \& {Mewes}}]{dAscoli2018ApJ...865..140D}
{d'Ascoli}, S., {Noble}, S.~C., {Bowen}, D.~B., {et~al.} 2018, \apj, 865, 140, \dodoi{10.3847/1538-4357/aad8b4}

\bibitem[{{D'Orazio} {et~al.}(2013){D'Orazio}, {Haiman}, \& {MacFadyen}}]{DOrazio2013MNRAS.436.2997D}
{D'Orazio}, D.~J., {Haiman}, Z., \& {MacFadyen}, A. 2013, \mnras, 436, 2997, \dodoi{10.1093/mnras/stt1787}

\bibitem[{{Ennoggi} {et~al.}(2026){Ennoggi}, {Campanelli}, {Krolik}, {Noble}, {Zlochower}, \& {de Simone}}]{Ennoggi2026PhRvL.136k1401E}
{Ennoggi}, L., {Campanelli}, M., {Krolik}, J., {et~al.} 2026, \prl, 136, 111401, \dodoi{10.1103/f74l-3c4y}

\bibitem[{{Event Horizon Telescope Collaboration} {et~al.}(2019){Event Horizon Telescope Collaboration}, {Akiyama}, {Alberdi}, {Alef}, {Asada}, {Azulay}, {Baczko}, {Ball}, {Balokovi{\'c}}, {Barrett}, {Bintley}, {Blackburn}, {Boland}, {Bouman}, {Bower}, {Bremer}, {Brinkerink}, {Brissenden}, {Britzen}, {Broderick}, {Broguiere}, {Bronzwaer}, {Byun}, {Carlstrom}, {Chael}, {Chan}, {Chatterjee}, {Chatterjee}, {Chen}, {Chen}, {Cho}, {Christian}, {Conway}, {Cordes}, {Crew}, {Cui}, {Davelaar}, {De Laurentis}, {Deane}, {Dempsey}, {Desvignes}, {Dexter}, {Doeleman}, {Eatough}, {Falcke}, {Fish}, {Fomalont}, {Fraga-Encinas}, {Friberg}, {Fromm}, {G{\'o}mez}, {Galison}, {Gammie}, {Garc{\'\i}a}, {Gentaz}, {Georgiev}, {Goddi}, {Gold}, {Gu}, {Gurwell}, {Hada}, {Hecht}, {Hesper}, {Ho}, {Ho}, {Honma}, {Huang}, {Huang}, {Hughes}, {Ikeda}, {Inoue}, {Issaoun}, {James}, {Jannuzi}, {Janssen}, {Jeter}, {Jiang}, {Johnson}, {Jorstad}, {Jung}, {Karami}, {Karuppusamy}, {Kawashima}, {Keating}, {Kettenis}, {Kim}, {Kim}, {Kim}, {Kino},
  {Koay}, {Koch}, {Koyama}, {Kramer}, {Kramer}, {Krichbaum}, {Kuo}, {Lauer}, {Lee}, {Li}, {Li}, {Lindqvist}, {Liu}, {Liuzzo}, {Lo}, {Lobanov}, {Loinard}, {Lonsdale}, {Lu}, {MacDonald}, {Mao}, {Markoff}, {Marrone}, {Marscher}, {Mart{\'\i}-Vidal}, {Matsushita}, {Matthews}, {Medeiros}, {Menten}, {Mizuno}, {Mizuno}, {Moran}, {Moriyama}, {Moscibrodzka}, {Mul{\ensuremath{\ddot{}}}ler}, {Nagai}, {Nagar}, {Nakamura}, {Narayan}, {Narayanan}, {Natarajan}, {Neri}, {Ni}, {Noutsos}, {Okino}, {Olivares}, {Oyama}, {{\"O}zel}, {Palumbo}, {Patel}, {Pen}, {Pesce}, {Pi{\'e}tu}, {Plambeck}, {PopStefanija}, {Porth}, {Prather}, {Preciado-L{\'o}pez}, {Psaltis}, {Pu}, {Ramakrishnan}, {Rao}, {Rawlings}, {Raymond}, {Rezzolla}, {Ripperda}, {Roelofs}, {Rogers}, {Ros}, {Rose}, {Roshanineshat}, {Rottmann}, {Roy}, {Ruszczyk}, {Ryan}, {Rygl}, {S{\'a}nchez}, {S{\'a}nchez-Arguelles}, {Sasada}, {Savolainen}, {Schloerb}, {Schuster}, {Shao}, {Shen}, {Small}, {Sohn}, {SooHoo}, {Tazaki}, {Tiede}, {Tilanus}, {Titus}, {Toma}, {Torne}, {Trent},
  {Trippe}, {Tsuda}, {van Bemmel}, {van Langevelde}, {van Rossum}, {Wagner}, {Wardle}, {Weintroub}, {Wex}, {Wharton}, {Wielgus}, {Wong}, {Wu}, {Young}, {Young}, {Younsi}, \& {Yuan}}]{EHT2019ApJ...875L...5E}
{Event Horizon Telescope Collaboration}, {Akiyama}, K., {Alberdi}, A., {et~al.} 2019, \apjl, 875, L5, \dodoi{10.3847/2041-8213/ab0f43}

\bibitem[{{Event Horizon Telescope Collaboration} {et~al.}(2022){Event Horizon Telescope Collaboration}, {Akiyama}, {Alberdi}, {Alef}, {Algaba}, {Anantua}, {Asada}, {Azulay}, {Bach}, {Baczko}, {Ball}, {Balokovi{\'c}}, {Barrett}, {Baub{\"o}ck}, {Benson}, {Bintley}, {Blackburn}, {Blundell}, {Bouman}, {Bower}, {Boyce}, {Bremer}, {Brinkerink}, {Brissenden}, {Britzen}, {Broderick}, {Broguiere}, {Bronzwaer}, {Bustamante}, {Byun}, {Carlstrom}, {Ceccobello}, {Chael}, {Chan}, {Chatterjee}, {Chatterjee}, {Chen}, {Chen}, {Cheng}, {Cho}, {Christian}, {Conroy}, {Conway}, {Cordes}, {Crawford}, {Crew}, {Cruz-Osorio}, {Cui}, {Davelaar}, {De Laurentis}, {Deane}, {Dempsey}, {Desvignes}, {Dexter}, {Dhruv}, {Doeleman}, {Dougal}, {Dzib}, {Eatough}, {Emami}, {Falcke}, {Farah}, {Fish}, {Fomalont}, {Ford}, {Fraga-Encinas}, {Freeman}, {Friberg}, {Fromm}, {Fuentes}, {Galison}, {Gammie}, {Garc{\'\i}a}, {Gentaz}, {Georgiev}, {Goddi}, {Gold}, {G{\'o}mez-Ruiz}, {G{\'o}mez}, {Gu}, {Gurwell}, {Hada}, {Haggard}, {Haworth}, {Hecht}, {Hesper},
  {Heumann}, {Ho}, {Ho}, {Honma}, {Huang}, {Huang}, {Hughes}, {Ikeda}, {Violette Impellizzeri}, {Inoue}, {Issaoun}, {James}, {Jannuzi}, {Janssen}, {Jeter}, {Jiang}, {Jim{\'e}nez-Rosales}, {Johnson}, {Jorstad}, {Joshi}, {Jung}, {Karami}, {Karuppusamy}, {Kawashima}, {Keating}, {Kettenis}, {Kim}, {Kim}, {Kim}, {Kim}, {Kino}, {Koay}, {Kocherlakota}, {Kofuji}, {Koch}, {Koyama}, {Kramer}, {Kramer}, {Krichbaum}, {Kuo}, {La Bella}, {Lauer}, {Lee}, {Lee}, {Leung}, {Levis}, {Li}, {Lico}, {Lindahl}, {Lindqvist}, {Lisakov}, {Liu}, {Liu}, {Liuzzo}, {Lo}, {Lobanov}, {Loinard}, {Lonsdale}, {Lu}, {Mao}, {Marchili}, {Markoff}, {Marrone}, {Marscher}, {Mart{\'\i}-Vidal}, {Matsushita}, {Matthews}, {Medeiros}, {Menten}, {Michalik}, {Mizuno}, {Mizuno}, {Moran}, {Moriyama}, {Moscibrodzka}, {M{\"u}ller}, {Mus}, {Musoke}, {Myserlis}, {Nadolski}, {Nagai}, {Nagar}, {Nakamura}, {Narayan}, {Narayanan}, {Natarajan}, {Nathanail}, {Navarro Fuentes}, {Neilsen}, {Neri}, {Ni}, {Noutsos}, {Nowak}, {Oh}, {Okino}, {Olivares}, {Ortiz-Le{\'o}n},
  {Oyama}, {{\"O}zel}, {Palumbo}, {Filippos Paraschos}, {Park}, {Parsons}, {Patel}, {Pen}, {Pesce}, {Pi{\'e}tu}, {Plambeck}, {PopStefanija}, {Porth}, {P{\"o}tzl}, {Prather}, {Preciado-L{\'o}pez}, \& {Psaltis}}]{EHT2022ApJ...930L..16E}
---. 2022, \apjl, 930, L16, \dodoi{10.3847/2041-8213/ac6672}

\bibitem[{{Farris} {et~al.}(2014){Farris}, {Duffell}, {MacFadyen}, \& {Haiman}}]{Farris2014ApJ...783..134F}
{Farris}, B.~D., {Duffell}, P., {MacFadyen}, A.~I., \& {Haiman}, Z. 2014, \apj, 783, 134, \dodoi{10.1088/0004-637X/783/2/134}

\bibitem[{{Farris} {et~al.}(2015){Farris}, {Duffell}, {MacFadyen}, \& {Haiman}}]{Farris2015MNRAS.446L..36F}
---. 2015, \mnras, 446, L36, \dodoi{10.1093/mnrasl/slu160}

\bibitem[{{Gold} {et~al.}(2014){Gold}, {Paschalidis}, {Etienne}, {Shapiro}, \& {Pfeiffer}}]{Gold2014PhRvD..89f4060G}
{Gold}, R., {Paschalidis}, V., {Etienne}, Z.~B., {Shapiro}, S.~L., \& {Pfeiffer}, H.~P. 2014, \prd, 89, 064060, \dodoi{10.1103/PhysRevD.89.064060}

\bibitem[{{Graham} {et~al.}(2015){Graham}, {Djorgovski}, {Stern}, {Glikman}, {Drake}, {Mahabal}, {Donalek}, {Larson}, \& {Christensen}}]{Graham2015Natur.518...74G}
{Graham}, M.~J., {Djorgovski}, S.~G., {Stern}, D., {et~al.} 2015, \nat, 518, 74, \dodoi{10.1038/nature14143}

\bibitem[{{Groselj} {et~al.}(2026){Groselj}, {Philippov}, {Beloborodov}, \& {Mushotzky}}]{Groselj2026arXiv260100518G}
{Groselj}, D., {Philippov}, A., {Beloborodov}, A.~M., \& {Mushotzky}, R. 2026, arXiv e-prints, arXiv:2601.00518, \dodoi{10.48550/arXiv.2601.00518}

\bibitem[{{Gr{\"u}nwald} {et~al.}(2023){Gr{\"u}nwald}, {Boller}, {Rakshit}, {Buchner}, {Dauser}, {Freyberg}, {Liu}, {Salvato}, \& {Schichtel}}]{Grunwald2023A&A...669A..37G}
{Gr{\"u}nwald}, G., {Boller}, T., {Rakshit}, S., {et~al.} 2023, \aap, 669, A37, \dodoi{10.1051/0004-6361/202244620}

\bibitem[{{Guti{\'e}rrez} {et~al.}(2022){Guti{\'e}rrez}, {Combi}, {Noble}, {Campanelli}, {Krolik}, {L{\'o}pez Armengol}, \& {Garc{\'\i}a}}]{Gutierrez2022ApJ...928..137G}
{Guti{\'e}rrez}, E.~M., {Combi}, L., {Noble}, S.~C., {et~al.} 2022, \apj, 928, 137, \dodoi{10.3847/1538-4357/ac56de}

\bibitem[{{Hirata} {et~al.}(1987){Hirata}, {Kajita}, {Koshiba}, {Nakahata}, {Oyama}, {Sato}, {Suzuki}, {Takita}, {Totsuka}, {Kifune}, {Suda}, {Takahashi}, {Tanimori}, {Miyano}, {Yamada}, {Beier}, {Feldscher}, {Kim}, {Mann}, {Newcomer}, {van}, {Zhang}, \& {Cortez}}]{Hirata1987PhRvL..58.1490H}
{Hirata}, K., {Kajita}, T., {Koshiba}, M., {et~al.} 1987, \prl, 58, 1490, \dodoi{10.1103/PhysRevLett.58.1490}

\bibitem[{{Hirose} {et~al.}(2006){Hirose}, {Krolik}, \& {Stone}}]{Hirose2006}
{Hirose}, S., {Krolik}, J.~H., \& {Stone}, J.~M. 2006, \apj, 640, 901, \dodoi{10.1086/499153}

\bibitem[{{Kallman} \& {Bautista}(2001)}]{Kallman2001ApJS..133..221K}
{Kallman}, T., \& {Bautista}, M. 2001, \apjs, 133, 221, \dodoi{10.1086/319184}

\bibitem[{{Kelly} {et~al.}(2009){Kelly}, {Bechtold}, \& {Siemiginowska}}]{Kelly2009ApJ...698..895K}
{Kelly}, B.~C., {Bechtold}, J., \& {Siemiginowska}, A. 2009, \apj, 698, 895, \dodoi{10.1088/0004-637X/698/1/895}

\bibitem[{Kinch {et~al.}(2020)Kinch, Noble, Schnittman, \& Krolik}]{Kinch_2020}
Kinch, B.~E., Noble, S.~C., Schnittman, J.~D., \& Krolik, J.~H. 2020, The Astrophysical Journal, 904, 117, \dodoi{10.3847/1538-4357/abc176}

\bibitem[{Kinch {et~al.}(2019)Kinch, Schnittman, Kallman, \& Krolik}]{Kinch_2019}
Kinch, B.~E., Schnittman, J.~D., Kallman, T.~R., \& Krolik, J.~H. 2019, The Astrophysical Journal, 873, 71, \dodoi{10.3847/1538-4357/ab05d5}

\bibitem[{Kinch {et~al.}(2021)Kinch, Schnittman, Noble, Kallman, \& Krolik}]{Kinch_2021}
Kinch, B.~E., Schnittman, J.~D., Noble, S.~C., Kallman, T.~R., \& Krolik, J.~H. 2021, The Astrophysical Journal, 922, 270, \dodoi{10.3847/1538-4357/ac2b9a}

\bibitem[{{Liao} {et~al.}(2021){Liao}, {Chen}, {Liu}, {Holgado}, {Guo}, {Gruendl}, {Morganson}, {Shen}, {Davis}, {Kessler}, {Martini}, {McMahon}, {Allam}, {Annis}, {Avila}, {Banerji}, {Bechtol}, {Bertin}, {Brooks}, {Buckley-Geer}, {Carnero Rosell}, {Carrasco Kind}, {Carretero}, {Javier Castander}, {Cunha}, {D'Andrea}, {da Costa}, {Davis}, {De Vicente}, {Desai}, {Thomas Diehl}, {Doel}, {Eifler}, {Evrard}, {Flaugher}, {Fosalba}, {Frieman}, {Garcia-Bellido}, {Gaztanaga}, {Glazebrook}, {Gruen}, {Gschwend}, {Gutierrez}, {Hartley}, {Hollowood}, {Honscheid}, {Hoyle}, {James}, {Krause}, {Kuehn}, {Lima}, {Maia}, {Marshall}, {Menanteau}, {Miquel}, {Plazas Malag{\'o}n}, {Roodman}, {Sanchez}, {Scarpine}, {Schubnell}, {Serrano}, {Smith}, {Smith}, {Soares-Santos}, {Sobreira}, {Suchyta}, {Swanson}, {Tarle}, {Vikram}, \& {Walker}}]{Liao2021MNRAS.500.4025L}
{Liao}, W.-T., {Chen}, Y.-C., {Liu}, X., {et~al.} 2021, \mnras, 500, 4025, \dodoi{10.1093/mnras/staa3055}

\bibitem[{{Liu} {et~al.}(2025){Liu}, {Nagele}, {Krolik}, {Kinch}, \& {Schnittman}}]{Liu+2025}
{Liu}, R., {Nagele}, C., {Krolik}, J.~H., {Kinch}, B.~E., \& {Schnittman}, J.~D. 2025, \apj, 982, 128, \dodoi{10.3847/1538-4357/adb61f}

\bibitem[{{Liu} {et~al.}(2016){Liu}, {Gezari}, {Burgett}, {Chambers}, {Draper}, {Hodapp}, {Huber}, {Kudritzki}, {Magnier}, {Metcalfe}, {Tonry}, {Wainscoat}, \& {Waters}}]{Liu2016ApJ...833....6L}
{Liu}, T., {Gezari}, S., {Burgett}, W., {et~al.} 2016, \apj, 833, 6, \dodoi{10.3847/0004-637X/833/1/6}

\bibitem[{{Liu} {et~al.}(2019){Liu}, {Gezari}, {Ayers}, {Burgett}, {Chambers}, {Hodapp}, {Huber}, {Kudritzki}, {Metcalfe}, {Tonry}, {Wainscoat}, \& {Waters}}]{Liu2019ApJ...884...36L}
{Liu}, T., {Gezari}, S., {Ayers}, M., {et~al.} 2019, \apj, 884, 36, \dodoi{10.3847/1538-4357/ab40cb}

\bibitem[{{Luo} {et~al.}(2025){Luo}, {Jiang}, \& {Liu}}]{Luo2025ApJ...978...86L}
{Luo}, D., {Jiang}, N., \& {Liu}, X. 2025, \apj, 978, 86, \dodoi{10.3847/1538-4357/ad9245}

\bibitem[{{MacFadyen} \& {Milosavljevi{\'c}}(2008)}]{MacFadyen2008ApJ...672...83M}
{MacFadyen}, A.~I., \& {Milosavljevi{\'c}}, M. 2008, \apj, 672, 83, \dodoi{10.1086/523869}

\bibitem[{{MacLeod} {et~al.}(2010){MacLeod}, {Ivezi{\'c}}, {Kochanek}, {Koz{\l}owski}, {Kelly}, {Bullock}, {Kimball}, {Sesar}, {Westman}, {Brooks}, {Gibson}, {Becker}, \& {de Vries}}]{MacLeod2010ApJ...721.1014M}
{MacLeod}, C.~L., {Ivezi{\'c}}, {\v{Z}}., {Kochanek}, C.~S., {et~al.} 2010, \apj, 721, 1014, \dodoi{10.1088/0004-637X/721/2/1014}

\bibitem[{{Malewicz} {et~al.}(2025){Malewicz}, {Ballantyne}, {Bogdanovi{\'c}}, {Brenneman}, \& {Dauser}}]{Malewicz2025ApJ...989..190M}
{Malewicz}, J., {Ballantyne}, D.~R., {Bogdanovi{\'c}}, T., {Brenneman}, L., \& {Dauser}, T. 2025, \apj, 989, 190, \dodoi{10.3847/1538-4357/adea75}

\bibitem[{{Malewicz} {et~al.}(2026){Malewicz}, {Bogdanovi{\'c}}, {Ballantyne}, {Brenneman}, \& {Dauser}}]{Malewicz2026arXiv260804961M}
{Malewicz}, J., {Bogdanovi{\'c}}, T., {Ballantyne}, D.~R., {Brenneman}, L., \& {Dauser}, T. 2026, arXiv e-prints, arXiv:2608.04961, \dodoi{10.48550/arXiv.2608.04961}

\bibitem[{{Manikantan} \& {Paschalidis}(2025)}]{Manikantan2025PhRvD.112j3050M}
{Manikantan}, V., \& {Paschalidis}, V. 2025, \prd, 112, 103050, \dodoi{10.1103/g98b-m8m1}

\bibitem[{{Manikantan} {et~al.}(2025){Manikantan}, {Paschalidis}, \& {Bozzola}}]{Manikantan2025PhRvD.112d3004M}
{Manikantan}, V., {Paschalidis}, V., \& {Bozzola}, G. 2025, \prd, 112, 043004, \dodoi{10.1103/js22-3hfx}

\bibitem[{{Marszewski} {et~al.}(2021){Marszewski}, {Prather}, {Joshi}, {Pandya}, \& {Gammie}}]{Marszewski2021ApJ...921...17M}
{Marszewski}, A., {Prather}, B.~S., {Joshi}, A.~V., {Pandya}, A., \& {Gammie}, C.~F. 2021, \apj, 921, 17, \dodoi{10.3847/1538-4357/ac1b28}

\bibitem[{{McHardy} {et~al.}(2004){McHardy}, {Papadakis}, {Uttley}, {Page}, \& {Mason}}]{McHardy2004MNRAS.348..783M}
{McHardy}, I.~M., {Papadakis}, I.~E., {Uttley}, P., {Page}, M.~J., \& {Mason}, K.~O. 2004, \mnras, 348, 783, \dodoi{10.1111/j.1365-2966.2004.07376.x}

\bibitem[{{Mihalas}(1985)}]{Mihalas1985JCoPh..57....1M}
{Mihalas}, D. 1985, Journal of Computational Physics, 57, 1, \dodoi{10.1016/0021-9991(85)90050-6}

\bibitem[{{Most} \& {Wang}(2024)}]{Most2024ApJ...973L..19M}
{Most}, E.~R., \& {Wang}, H.-Y. 2024, \apjl, 973, L19, \dodoi{10.3847/2041-8213/ad7713}

\bibitem[{{Most} \& {Wang}(2025)}]{Most2025PhRvD.111h1304M}
---. 2025, \prd, 111, L081304, \dodoi{10.1103/PhysRevD.111.L081304}

\bibitem[{{Mundim} {et~al.}(2014){Mundim}, {Nakano}, {Yunes}, {Campanelli}, {Noble}, \& {Zlochower}}]{Mundim2014PhRvD..89h4008M}
{Mundim}, B.~C., {Nakano}, H., {Yunes}, N., {et~al.} 2014, \prd, 89, 084008, \dodoi{10.1103/PhysRevD.89.084008}

\bibitem[{{Nagele} {et~al.}(2026{\natexlab{a}}){Nagele}, {Krolik}, {Kinch}, \& {Schnittman}}]{Nagele+2026b}
{Nagele}, C., {Krolik}, J.~H., {Kinch}, B.~E., \& {Schnittman}, J.~D. 2026{\natexlab{a}}, \apj, 1005, 119, \dodoi{10.3847/1538-4357/ae6eff}

\bibitem[{{Nagele} {et~al.}(2026{\natexlab{b}}){Nagele}, {Krolik}, {Liu}, {Kinch}, \& {Schnittman}}]{Nagele+2026a}
{Nagele}, C., {Krolik}, J.~H., {Liu}, R., {Kinch}, B.~E., \& {Schnittman}, J.~D. 2026{\natexlab{b}}, \apj, 1001, 190, \dodoi{10.3847/1538-4357/ae4ec7}

\bibitem[{{Noble} \& {Krolik}(2009)}]{Noble2009ApJ...703..964N}
{Noble}, S.~C., \& {Krolik}, J.~H. 2009, \apj, 703, 964, \dodoi{10.1088/0004-637X/703/1/964}

\bibitem[{{Noble} {et~al.}(2021){Noble}, {Krolik}, {Campanelli}, {Zlochower}, {Mundim}, {Nakano}, \& {Zilh{\~a}o}}]{Noble2021ApJ...922..175N}
{Noble}, S.~C., {Krolik}, J.~H., {Campanelli}, M., {et~al.} 2021, \apj, 922, 175, \dodoi{10.3847/1538-4357/ac2229}

\bibitem[{{Noble} {et~al.}(2009){Noble}, {Krolik}, \& {Hawley}}]{Noble_2009}
{Noble}, S.~C., {Krolik}, J.~H., \& {Hawley}, J.~F. 2009, \apj, 692, 411, \dodoi{10.1088/0004-637X/692/1/411}

\bibitem[{{Noble} {et~al.}(2012){Noble}, {Mundim}, {Nakano}, {Krolik}, {Campanelli}, {Zlochower}, \& {Yunes}}]{Noble2012ApJ...755...51N}
{Noble}, S.~C., {Mundim}, B.~C., {Nakano}, H., {et~al.} 2012, \apj, 755, 51, \dodoi{10.1088/0004-637X/755/1/51}

\bibitem[{{Novikov} \& {Thorne}(1973)}]{Novikov1973blho.conf..343N}
{Novikov}, I.~D., \& {Thorne}, K.~S. 1973, in Black Holes (Les Astres Occlus), ed. C.~{Dewitt} \& B.~S. {Dewitt}, 343--450

\bibitem[{{Pandya} {et~al.}(2016){Pandya}, {Zhang}, {Chandra}, \& {Gammie}}]{Pandya2016ApJ...822...34P}
{Pandya}, A., {Zhang}, Z., {Chandra}, M., \& {Gammie}, C.~F. 2016, \apj, 822, 34, \dodoi{10.3847/0004-637X/822/1/34}

\bibitem[{{Paschalidis} {et~al.}(2021){Paschalidis}, {Bright}, {Ruiz}, \& {Gold}}]{Paschalidis2021ApJ...910L..26P}
{Paschalidis}, V., {Bright}, J., {Ruiz}, M., \& {Gold}, R. 2021, \apjl, 910, L26, \dodoi{10.3847/2041-8213/abee21}

\bibitem[{{Reynolds} \& {Miller}(2009)}]{Reynolds2009ApJ...692..869R}
{Reynolds}, C.~S., \& {Miller}, M.~C. 2009, \apj, 692, 869, \dodoi{10.1088/0004-637X/692/1/869}

\bibitem[{{Roedig} {et~al.}(2014){Roedig}, {Krolik}, \& {Miller}}]{Roedig2014ApJ...785..115R}
{Roedig}, C., {Krolik}, J.~H., \& {Miller}, M.~C. 2014, \apj, 785, 115, \dodoi{10.1088/0004-637X/785/2/115}

\bibitem[{{Schnittman} \& {Krolik}(2013)}]{Schnittman_2013b}
{Schnittman}, J.~D., \& {Krolik}, J.~H. 2013, \apj, 777, 11, \dodoi{10.1088/0004-637X/777/1/11}

\bibitem[{Schnittman {et~al.}(2013)Schnittman, Krolik, \& Noble}]{Schnittman_2013}
Schnittman, J.~D., Krolik, J.~H., \& Noble, S.~C. 2013, The Astrophysical Journal, 769, 156, \dodoi{10.1088/0004-637X/769/2/156}

\bibitem[{{Shakura} \& {Sunyaev}(1973)}]{Shakura1973A&A....24..337S}
{Shakura}, N.~I., \& {Sunyaev}, R.~A. 1973, \aap, 24, 337

\bibitem[{{Shi} {et~al.}(2012){Shi}, {Krolik}, {Lubow}, \& {Hawley}}]{Shi2012ApJ...749..118S}
{Shi}, J.-M., {Krolik}, J.~H., {Lubow}, S.~H., \& {Hawley}, J.~F. 2012, \apj, 749, 118, \dodoi{10.1088/0004-637X/749/2/118}

\bibitem[{{Tang} {et~al.}(2018){Tang}, {Haiman}, \& {MacFadyen}}]{Tang2018MNRAS.476.2249T}
{Tang}, Y., {Haiman}, Z., \& {MacFadyen}, A. 2018, \mnras, 476, 2249, \dodoi{10.1093/mnras/sty423}

\bibitem[{{Tiede} {et~al.}(2020){Tiede}, {Zrake}, {MacFadyen}, \& {Haiman}}]{Tiede2020ApJ...900...43T}
{Tiede}, C., {Zrake}, J., {MacFadyen}, A., \& {Haiman}, Z. 2020, \apj, 900, 43, \dodoi{10.3847/1538-4357/aba432}

\bibitem[{{Tiwari} {et~al.}(2025){Tiwari}, {Chan}, {Bogdanovi{\'c}}, {Jiang}, {Davis}, \& {Ferrel}}]{Tiwari2025ApJ...986..158T}
{Tiwari}, V., {Chan}, C.-H., {Bogdanovi{\'c}}, T., {et~al.} 2025, \apj, 986, 158, \dodoi{10.3847/1538-4357/add408}

\bibitem[{{Vaughan} {et~al.}(2016){Vaughan}, {Uttley}, {Markowitz}, {Huppenkothen}, {Middleton}, {Alston}, {Scargle}, \& {Farr}}]{Vaughan2016MNRAS.461.3145V}
{Vaughan}, S., {Uttley}, P., {Markowitz}, A.~G., {et~al.} 2016, \mnras, 461, 3145, \dodoi{10.1093/mnras/stw1412}

\bibitem[{{Veronesi} {et~al.}(2026){Veronesi}, {Charisi}, {Petrov}, {Taylor}, {Runnoe}, {D'Orazio}, {Pilawa}, \& {Ma}}]{Veronesi2026arXiv260600218V}
{Veronesi}, N., {Charisi}, M., {Petrov}, P., {et~al.} 2026, arXiv e-prints, arXiv:2606.00218, \dodoi{10.48550/arXiv.2606.00218}

\bibitem[{{Zilh{\~a}o} {et~al.}(2015){Zilh{\~a}o}, {Noble}, {Campanelli}, \& {Zlochower}}]{Zilhao2015PhRvD..91b4034Z}
{Zilh{\~a}o}, M., {Noble}, S.~C., {Campanelli}, M., \& {Zlochower}, Y. 2015, \prd, 91, 024034, \dodoi{10.1103/PhysRevD.91.024034}

\bibitem[{{Zrake} {et~al.}(2021){Zrake}, {Tiede}, {MacFadyen}, \& {Haiman}}]{Zrake2021ApJ...909L..13Z}
{Zrake}, J., {Tiede}, C., {MacFadyen}, A., \& {Haiman}, Z. 2021, \apjl, 909, L13, \dodoi{10.3847/2041-8213/abdd1c}

\end{thebibliography}
\bibliographystyle{aasjournal}

\end{document}